\documentclass[12pt]{amsart}

\usepackage[margin=1in]{geometry}
\usepackage[lofdepth,lotdepth,caption=false]{subfig}
\usepackage{fancyhdr}
\usepackage{hyperref}
\usepackage{faktor}
\usepackage{amsmath, amssymb, graphicx}
\usepackage{mathtools}
\usepackage{xspace}
\usepackage{braket}
\usepackage{yfonts}
\usepackage{multicol}
\usepackage{multirow}
\usepackage{color}
\usepackage{setspace}
\usepackage{enumitem}
\usepackage{pst-node}
\usepackage{tikz-cd}
\usepackage{tikz}
\usepackage{tikz-network}
\usetikzlibrary{decorations.markings}
\usepackage{slashed}
\usetikzlibrary{calc}
\usepackage[mathscr]{euscript}
\usepackage[numbers]{natbib}
\newcommand{\ra}{\rightarrow}
\newcommand{\pr}{\prime}

\newcommand{\C}{\mathbb{C}}

\newcommand{\R}{\mathbb{R}}

\newcommand{\abs}[1]{\left\lvert #1 \right\rvert}
\newcommand{\tld}[1]{\widetilde{#1}}

\newcommand{\id}{\mathrm{id}}

\DeclareMathAlphabet{\mathpzc}{OT1}{pzc}{m}{it}

\newcommand{\mcD}{\mathcal{D}}

\newcommand{\mcH}{\mathcal{H}}

\newcommand{\mcM}{\mathcal{M}}

\newcommand{\mcZ}{\mathcal{Z}}

\theoremstyle{plain}  

\newtheorem{thm}{Theorem}[section]

\newtheorem{cor}[thm]{Corollary}

\theoremstyle{definition}
\newtheorem{dfn}{Definition}[section]
\theoremstyle{remark}
\newtheorem{rmk}{Remark}[section]
\newtheorem{fct}{Fact}[section]

\begin{document}

\title{The Universe from a Single Particle}
\maketitle
\begin{center}
  \normalsize
  Michael Freedman \footnote{michaelf@microsoft.com}\textsuperscript{,\hyperref[1]{*}},
  Modjtaba Shokrian Zini \footnote{mshokrianzini@pitp.ca; Research was done as part of an internship of this author at Microsoft} \textsuperscript{,\hyperref[2]{$\dagger$}}
 \par   \bigskip
\end{center}
\begin{abstract}
    We explore the emergence of many-body physics from quantum mechanics via spontaneous symmetry breaking. To this end, we study potentials which are functionals on the space of Hamiltonians enjoying an unstable critical point corresponding to a random quantum mechanical system (the Gaussian unitary ensemble), but also less symmetrical local minima corresponding to interacting systems at the level of operators.
\end{abstract}

\tableofcontents
\section{Introduction}
Since Wilson \cite{RevModPhys.47.773} much attention both in condensed matter and high energy has been devoted to the question: Can theory B emerge as a low energy limit of theory A? And even possible concatenations $\text{A} \ra \text{B} \ra \text{C} \ra \ldots$ ? Effective field theories are undoubtedly important links in such chains. This paper explores, in a toy model, the possibility that field theory itself, even at high energy, may arise through a kind of symmetry breaking applied to a much simpler system, single particle quantum mechanics.

In our toy model, single particle quantum mechanics is represented by theory A with a finite dimensional Hilbert space of states $\mcH_A \cong \C^N$. Initially, and for simplicity, we take $N$ to be a power of 2, $N = 2^n$, but, as we will see, this choice simplifies the discussion but may not be essential. We study symmetry breaking to a theory B which is a toy model for field theory. Theory B has a Hilbert space $\mcH_B$ isometric to $\underbrace{\C^2 \otimes \cdots \otimes \C^2}_{n \text{ times}}$ but $\mcH_B$ is regarded as describing a system of $n$ interacting bosons, each with two states (a very similar choice, not explored here, is for $\mcH_B$ to describe $n$ interacting spin-$\frac{1}{2}$ Fermions). The difference between models A and B is the probability distribution from which the system Hamiltonian, which we think of as the Hamiltonian of the universe, is drawn. In case A, we consider that the Hamiltonian $H_A$ is drawn \emph{randomly} from the Gaussian Unitary Ensemble (GUE) $\mcZ$ on $\mathfrak{su}(N)$, the Lie algebra of the group of symmetries of $\mcH_A$. We consider the simple Lie group $\operatorname{SU}(N)$ to be the symmetry group of $\mcH_A$ as overall phase is unobservable. Since the elements of $\mathfrak{su}(N)$ are skew-Hermitian, we multiply by $-i$ to get a Hermitian Hamiltonian. To complete the definition of theory A we specify the metric $g_{ij}$ on $\mathfrak{su}(N)$. We choose $g_{ij}$ to be the (ad-invariant) killing form: $\langle a, b \rangle \propto \operatorname{tr}(\operatorname{ad}(a) \circ \operatorname{ad}(b))$. Specifying the metric makes the distribution $\mcZ$ well-defined and induces a measure on $\mathfrak{su}(N)$, giving meaning to the word ``randomly'' above. In the case $g_{ij}$ is the Killing form, the resulting distribution is the GUE.

Concretely, imagine a particle with access to $N$ states $\ket{1},\ldots,\ket{N}$, with the system Hamiltonian $H_A$ built by assembling $N^2$ i.i.d real Gaussian random variables into
\begin{enumerate}
    \item the diagonal entries of $H_A$,
    \item and the real and imaginary parts of $(H_A)_{ij}, i>j$ with $(H_A)_{ij} = \overline{(H_A)_{ji}}$.
\end{enumerate} 
Constructing $H_A$ in this way reflects a complete agnosticism on the possible transitions, making $H_A$ the Hamiltonian for a random single particle system.

By contrast, in case of B, some qubit structure has been chosen in the form of an isometry: $\C^{2^n} \overset{J}{\cong} \underbrace{\C^2 \otimes \cdots \otimes \C^2}_n$. Thus the moduli space of qubits structures is the homogeneous space 
$$\textswab{M}  = \faktor{U(2^n)}{\underbrace{U(2)\times \cdots \times U(2)}_{n}},$$
where $J$ determines the inclusion. There is now a central definition:

\begin{dfn}\label{dfnkaq}
A non-singular inner product (i.e. a metric) $g_{ij}$ on $\mathfrak{su}(2^n)$ which is \textbf{not} bi-invariant, \textbf{k}nows \textbf{a}bout \textbf{q}ubits (is \textbf{kaq}) if there is a basis $\{iH_1, \ldots, iH_{4^n-1} \}$ for $\mathfrak{su}(2^n)$ consisting of principal axes for $g_{ij}$ with the property that there is an isomorphism of Lie algebras
$$j: \underbrace{\mathfrak{u}(2)\otimes \cdots \otimes \mathfrak{u}(2)}_{n \text{ copies}} \to \mathfrak{u}(2^n)$$
such that $iH_k = j(O_{1,k} \otimes \ldots \otimes O_{n,k})$ for $1 \le k \le 4^n -1$, where $O_{i,k}$ lives in the $i$-th copy of $\mathfrak{u}(2)$; that is, is there is a basis of principal axes which are \textit{pure tensors} relative to $j$. 
\end{dfn}
The previous choice $J$ determines $j$. The notions of isometry and principal axis above are with respect to a unique (up to scale) bi-invariant metric $\left< a,b\right> = \frac{1}{\dim \mathfrak{su}(N)} \operatorname{tr}(\operatorname{ad}(a) \circ \operatorname{ad}(b))$. Implicitly, this definition exploits the natural inclusion $\operatorname{SU}(N) \subset \operatorname{U}(N)$, since our $g_{ij}$ is defined only on the former.

For clarity, we remark that if the (inverse) bi-invariant metric is used to raise an index of $g_{ij}$ to create a symmetric operator $g_i^j$, then the principal axes of $g_{ij}$ are precisely the eigenvectors of $g_i^j$. The reason the definition refers to \textbf{a} basis of principal axes (eigenvectors) instead of \textbf{the} basis is that in the metrics we encounter there is usually degeneracy (to numerical precision) so we need the freedom to look within eigenspaces for a basis of \textit{simple} (pure tensor form) eigenvectors. In practice, one might wish to soften the definition slightly and allow eigenspaces with \textit{nearly} identical eigenvalues also to be combined, thus allowing more freedom to find simple linear combinations. 

We can generalize the definition above to \textbf{k}nows-\textbf{a}bout-\textbf{q}u\textbf{n}its (\textbf{kaqn}). We have used qu\textbf{b}its to define \textbf{kaq} but the reader can imagine other primes or combination of primes not restricted to 2. In this initial paper, we have not attempted computations in this direction but expect to study this in future works.

One challenge we face is the large dimension $4^n-3n-1$ of $\textswab{M}$. To decide if a given metric $g_{ij}$ on $\mathfrak{su}(2^n)$ is or is not \textbf{kaq}, it may be necessary to numerically search $\textswab{M}$. For $n=2$, we did this on several occasions. But a particularly simple family of \textbf{kaq} metrics are those where there is a basis of principal axes of $g_{ij}$ consisting of Pauli words. The structure $J$ determines a \emph{Pauli-word} basis, called $\textbf{PB}_n$ on the Lie algebra $\mathfrak{su}(2^n) = $
\[
    \{-i X \otimes 1 \otimes \cdots 1, i Y \otimes 1 \otimes \cdots \otimes 1, -i Z \otimes 1 \otimes \cdots \otimes 1, -i X \otimes Y \otimes 1 \cdots \otimes 1, \dots, -i  Z \otimes Z \otimes \cdots \otimes Z\},
\]
consisting of all words of length $n$ in the letters $\{1, X, Y, Z\}$ except the all 1 word which is in $\mathfrak{u}(2^n)$ but not $\mathfrak{su}(2^n)$. Given $J$, this basis is canonical up to the action of $\operatorname{SU}(2)$ on each factor. 

Hence forth, we normalize all metrics (also the killing form) so that $\operatorname{det}(g_{ij}) = 1$. A very interesting special case of \textbf{kaq} metrics on $\mathfrak{su}(2^n)$ are the previously studied \emph{penalty} metrics of \cite{nielsen2005geometric,nielsen2006quantum,dowling2008geometry}, \cite{brown2017quantum,brown2018second}, and \cite{BFLS} where
\begin{equation}\label{penalty_metric}
    g_{ij} \propto e^{r w(i)} \delta_{ij}
\end{equation}
where $r$ is some positive constant and $w(i)$ is the \emph{weight} of the $i$th Pauli-word meaning the number of letters it contains, X, Y, or Z, which are different from 1. Of course:
\begin{equation}
    X = \begin{pmatrix}
        0 & 1 \\ 1 & 0
    \end{pmatrix},\
    Y = \begin{pmatrix}
        0 & -i \\ i & 0
    \end{pmatrix},\
    Z = \begin{pmatrix}
        1 & 0 \\
        0 & -1
    \end{pmatrix}.
\end{equation}
We should emphasize that \textbf{kaq} metrics are exceedingly rare having roughly square root the number of parameters as general metrics. A \textbf{kaq} metric needs $4^n-1$ parameters to specify $j$ (on which $\operatorname{Aut}(\mathfrak{su}(2^n))$ acts freely and transitively), and also $4n$ parameters for each product of $O_{i,k}$'s of which there are $4^n-1$ (as $1\le k \le 4^n -1$). In total this makes $(4^n-1)+(4^n-1)(4n) = (4n+1)(4^n-1)$ parameters for a \textbf{kaq} metric, whereas the metrics space $g_{ij}$ on $\mathfrak{su}(2^n)$ has dimension $\frac{4^n(4^n-1)}{2}$.

This paper studies symmetry breaking from theory A to theory B. This is a symmetry breaking at the level of (probability distributions on) operators, not of states as is familiar, for example, in the Ginzburg-Landau theory of superconductivity. To do this, we consider (one of several possible) functionals:
\begin{equation}
    f: \{\text{metrics}\} \ra \R
\end{equation}
on the space of $\operatorname{det} = 1$ inner products (i.e.\ normalized metrics) on $\mathfrak{su}(2^n)$ and ask: Does $f$ break the symmetry of the bi-invariant Killing form to a \textbf{kaq} metric?

\begin{figure}[!ht]
    \centering
    \begin{tikzpicture}[scale=1.3]
        \draw [->] (-4,0) -- (4,0);
        \node at (4.7,0.05) {metrics};
        \draw [->] (4,0.3) -- (4,4);
        \node[rotate=90] at (4.25,2) {Real};
        \node at (0,0.3) {Killing form};
        \draw (-3.7,3.2) to [out=-90,in=180] (-2.6,1) to [out=0,in=180] (-2.1,1.3) to [out=0,in=180] (-1.6,1) to [out=0,in=180] (-1.2,1.15) to [out=0,in=180] (-0.8,1) to [out=0,in=180] (0,2) to[out=0,in=180] (0.8,1) to [out=0,in=180] (1.2,1.15) to [out=0,in=180] (1.6,1) to [out=0,in=180] (2.1,1.3) to [out=0,in=180] (2.6,1) to[out=0,in=-90] (3.7,3.2);
        \draw [->] (0,0.5) -- (0,1.8);
        \node at (-3.6,0.3) {other};
        \draw [->] (-3.25,0.4) -- (-2.9,0.9);
        \node at (3.1,0.3) {\textbf{kaq} metric};
        \draw [->] (2.3,0.4) -- (1.7,0.9);
        \node at (2.8,2.7) {graph $f$};
        \draw [->] (2.8,2.4) to[out=-90,in=180] (3.4,2);
    \end{tikzpicture}
    \caption{Conjectural symmetry breaking from Killing to \textbf{kaq}.}
    \label{mexicanhat}
\end{figure}
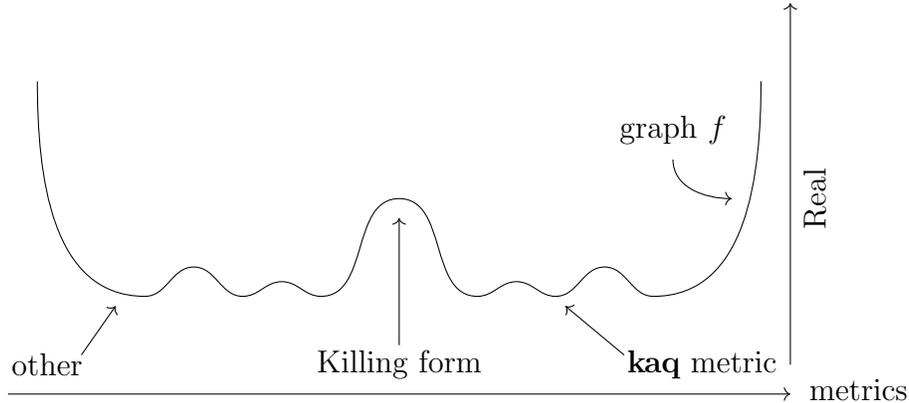

Let us repeat what we are looking for and why we are looking for it.

First the ``what''. We are looking for a functional $f$ on metrics which, if minimized, would confer the usual features of interacting physics on what, at a fundamental level, is actually a single particle system. The systems (A and B) we study are each fully quantum mechanical but are each drawn from their separate statistical ensembles: Gaussians defined from quite different inner products. Because the overwhelming preponderance of \textbf{kaq} metrics produce measures with infinitesimally small overlap with the ad-invariant measure (induced by the Killing form), a \textbf{kaq} metric will display observable features of many-body physics, almost certainly not seen in a Killing-random (GUE) Hamiltonian from theory A. The distinction is most clear in the special case of penalty metric in which the expected strength of $k$-body terms in the Hamiltonian $H_B$, such as
\begin{equation}
    1 1 \underbrace{X Y Z \dots Z Y}_{k \text{ letters}} 1 \dots 1
\end{equation}
($\otimes$ is suppressed and the non-identity letters are not required to be consecutive), will decay exponentially with $k$.

If one were handed a seemingly random $2^n \times 2^n$ Hermitian matrix and discovered (not an easy computational task) that it was mostly captured, in some tensor coordinates, by low weight terms, one would believe many-body physics was at work. The SYK model $H = \sum J_{ijkl} \gamma_i \gamma_j \gamma_k \gamma_l$ \cite{trunin2020pedagogical} is a (fermionic) illustration. It assembles the system Hamiltonian for a black hole from random 4-Majorana interactions (analogous to 2-body interactions in the bosonic case) governed by a complete graph (no ``spatial locality''). The SYK model can easily be enhanced, in the spirit of penalty metrics, to allow exponentially weakening higher body interactions:
\begin{equation}
    H = \sum_{p=2}^n e^{\text{const . }p}\left(\sum_{i_1, \ldots, i_{2p}} J_{i_1 \dots i_{2p}} \gamma_{i_1} \cdots \gamma_{i_{2p}}\right)
\end{equation}

Such Hamiltonians are part and parcel of many-body physics. It would indeed be surprising to be told that the system was ``really'' a single particle whose governing Hamiltonian had rolled off an unstable equilibrium into a \textbf{kaq} well. This is the surprise we are investigating.

Before addressing the ``why,'' a word about the parameter space $\{\text{metrics}\}$ and the functionals $f$ we investigate.

The geometry of compact simple Lie groups with bi-invariant metrics is very well understood \cite{helgason1979differential}. They are symmetric spaces whose geodesics are the cosets of 1-parameter subgroups. In this case much global information is known: for example, all points on the cut locus (say from $\id$) are also conjugate points. On the other hand, if the Riemannian metric is merely left-invariant the geometry is a rich subject of contemporary research. Local computations are straightforward \cite{milnor1976curvatures,berestovskii1988curvatures} but global properties are harder to derive (in \cite{BFLS} it was recently shown that for a fixed constant $c > 0$ the diameter of the penalty metrics on $\operatorname{SU}(2^n)$ is larger by a factor, exponential in $n$, than the Killing metric).

Choosing a left-invariant metric on a Lie group $G$ is of course the same thing as choosing an inner product on its Lie algebra $\mathfrak{g}$ (and moving that inner product about by the differential of left multiplication). So when we study $\{\text{metrics}\}$ on $\operatorname{su}(2^n)$ we are really studying left invariant Riemannian geometries on $\operatorname{SU}(2^n)$. The normalization condition $\operatorname{det}(g_{ij}) = 1$ ensures a fixed volume for $\operatorname{SU}(2^n)$ as the metric varies. The penalty metrics make high weight directions exponentially expensive so a typical $H_B$ for such a metric will consist mostly of low-weight (i.e. few-body) interactions, i.e.\ look like typical many-body physics. So we are looking for functionals which break symmetry to penalty metrics or at least \textbf{kaq} metrics. It is clearly \emph{cheating} to mention a qubit structure in the \emph{definition} of $f$. $f$ should be defined only from what is intrinsic to noninteracting physics, the structure constants of $\mathfrak{su}(2^n)$, $c_{ij}^{k}$, and the metric $g_{ij}$.

As geometers, a very natural functional might be $f = -$diameter. For the Killing metric, the diameter is $\pi$ but can be substantially increased by varying $g_{ij}$ from $\id$. While $f = -$diameter might eventually reward investigation, we must restrict ourselves to studying functionals $f$ where computations are efficient. So we have tried to assemble for study the most natural functionals built from the tensors $g_{ij}$ and $c_{ij}^{k}$. For $f$, we settled on Ricci scalar curvature which is easily expressed in terms of $g_{ij}$ and $c_{ij}^k$, as well as a variety of (cubicly) perturbed finite dimensional Gaussian integrals. The practical reason for studying these is that it is easy to work out the first few terms of their perturbative expansions in terms of tensor contraction diagrams. Once (a few terms of) the perturbative expansion is in hand, there are efficient methods, e.g.\ Pad\'{e} approximates \cite{bender2013advanced} to achieve \emph{rapid} convergence (to be explored in future works) but merely summing the first few terms is useful when the coupling $k$ to the cubic term is small. These integrals are described in the next section.

Regrading the accuracy of our methods, a mathematical admission is in order. We do not have a proof that the integrals we use to define our functional $f$ have an analytic meaning. This is currently under investigation with preliminary remarks in section 4. We proceed instead, as one often does in field theory, trusting that the integral expression, through some regularization, actually defines a function of a (complex) expansion parameter and trust that detailed information on this function can be extracted from its asymptotic expansion (possibly post-processed via Pad\'{e} approximates). Of course, this issue is not present when $f=$ Ricci scalar curvature.

Finally, why? The Wilsonian dream is that at highest energy the universe has a simple mathematical description and that any complexity in the apparent physical laws arose through spontaneous symmetry breaking (SSB). For example:

\begin{figure}[!ht]
    \centering
    \begin{tikzpicture}
        \node at (0,0) {boundary CFT $\xrightarrow{\text{dual}} \begin{pmatrix} \text{stringy} \\ \text{bulk theory} \end{pmatrix} \xrightarrow{\text{SSB}}\ ? \rightarrow \cdots \xrightarrow{\text{SSB}} \text{ Standard Model}$};
        \node at (4.7,-0.4) {$\cup$};
        \node at (4.7,-0.8) {Quantum mechanics (QM)};
        \node at (4.7,-1.2) {(single particle)};
    \end{tikzpicture}
    \caption{A typical Wilsonian dream.}
    \label{fig:wilson}
\end{figure}

All the theories in any such story have necessarily been interacting because they are intended to model an obviously interacting universe.  The goal of this paper is to propose, in the context of a highly restricted toy model, that a prequel is possible. That single particle quantum mechanics could also be the beginning of the story. Adding quantum mechanics also to the far left in Figure \ref{fig:wilson}, we propose that SSB, at the operator level, could allow QM to simulate the interacting systems from which the familiar story is told, as shown in Figure \ref{fig:wilsonqm}.

\begin{figure}[!ht]
    \centering
    \begin{tikzpicture}
        \node at (0,0) {$\begin{pmatrix} \text{QM on} \\ \text{Hilbert space} \\ \cong \C^N \end{pmatrix} \xrightarrow{\text{SSB}}$ boundary CFT $\xrightarrow{\text{dual}} \begin{pmatrix} \text{stringy} \\ \text{bulk theory} \end{pmatrix} \xrightarrow{\text{SSB}}\ ? \rightarrow \cdots \xrightarrow{\text{SSB}} \text{ Standard Model}$};
        \node at (6.7,-0.4) {$\cup$};
        \node at (6.7,-0.9) {QM};
    \end{tikzpicture}
    \caption{Dream with prequel.}
    \label{fig:wilsonqm}
\end{figure}

Although the vision is broad, the goal of this paper is modest: explore symmetry breaking from QM to a \textbf{kaq}-geometry. The reader will naturally wonder when we will make contact with experiment or observation. Our aim here, is quite low but fundamental: to address the fact that the universe appears to contain more than one thing. It would be nice to deduce: the dimension(s) of spacetime, the structure of physical laws, the existence of fermions, and the value of the fine-structure constant. But the truth is that we are nowhere near ready to address these questions. However, to the large ensemble of untouched questions, we would like to add a speculation.

It may have, and should have, seemed unnatural that for theory A, our random finite dimensional QM system was placed on a Hilbert space of dimension $N = 2^n$. What would one expect if $N$ was just a randomly selected large integer? Could there still be a symmetry breaking story to a collection of various qunits? If $N$ is chosen at random in a certain size range (say 100 digit numbers), we can ask what is the expected size of its largest prime factor. The answer is that it likely will have about 62 digits. This phenomenon is governed by the Golomb-Dickman constant $\approx 0.624329989 \dots$. Much beyond this is known regarding the statistics of prime factorizations of large random integers. If $N$ instead of being a power of 2, is random, perhaps a suitable functional $f$ will break the large system into qunits of dimensions the prime factors of $N$.

As a test of our symmetry breaking thesis, one should look near the r.h.s. of Figure \ref{fig:wilsonqm} for some phenomenological residue of a factorization of the initial random integer $N$. $N$ should indeed be large. Since the entropy of measured black holes can be $10^{90}$ \cite{PhysRevD.7.2333,egan2010larger} we would need $n >> 10^{90}$ and $N >> 2^{10^{90}}$, roughly what was once called Googolplex. To understand how the prime factorization of a large random integer might be imprinted into the low energy physics is a major challenge, and one approach to gathering physical evidence for the symmetry breaking hypothesis that we study here on a mathematical plane. The distribution of primes is tightly reflected in the Riemann zeta function, and similar zeta functions built from the length distribution of closed geodesics on a hyperbolic surface are linked through the Selberg trace formula \cite{selberg1956harmonic} to the spectrum of the Laplacian on these surfaces. So statistical properties of factorization could have an audible echo in string dynamics if we know what to listen for.

Another mathematical possibility to be explored is that the prime 2 is energetically favored because of the relation to Majorana/Clifford algebras, and that for a general integer $N$, SSB will simply round down to a power of 2, splitting the Hilbert space into a direct sum of glassy physics, and a left-over degrees of freedom. The size of the power of 2 may depend on energy/entropy balance, as there should be many more shallow, smaller Clifford algebra minima and fewer larger and deeper ones. Leftover degrees of freedom may separately organize themselves into additional interacting systems. A Hamiltonian which is a superposition of several separate interacting sectors would govern separate world branches which, like the live and dead cat, do not know about each other.

In future works, we hope to investigate SSB to ``partial''-\textbf{kaq}(\textbf{n}) minima in which a proper summand of the Hilbert space is \textbf{kaq}(\textbf{n}), or the possibility of ``multi''-\textbf{kaq}(\textbf{n}) minima where the Hilbert space decomposes into several subspaces each of which has a \textbf{kaq}(\textbf{n}) decomposition. For example, one might find a 17D Hilbert space decomposing as $(\mathbb{C}^3)^{\otimes 2} \oplus (\mathbb{C}^2)^{\otimes 3}$.

Let us finish this introductory section with integrals used to define functionals $f: \{\text{metrics}\} \ra \R$, and a discussion of their symmetries. A discussion of the numerics and ML used to enhance the numerics is in the next sections.

We define the complex-valued functions: $F_k: \{\text{metrics}\} \ra \C$, and consider functions of the form
\begin{align}
    & f_{k,1} = \operatorname{Re}(F_k),\ f_{k,2} = \operatorname{Im}(F_k),\ f_{k,3} = \abs{F_k}^2,\ k \in \C \\ \label{theintegral}
    & F_k = \int_{\vec{x} \in \R^{3(4^n-1)}}\ \operatorname{d}\vec{x}\ e^{ik(G_{IJ}x^I x^J + c_{ijk}y_1^i y_2^j y_3^k)} 
\end{align}
for $x = (y_1,y_2,y_3)$ with $y_o \in \R^{4^n-1}, o \in \{1,2,3\}$, and for $I = (i,o)$, $x^I = y_{o}^i \in \mathbb{R}$, and $G_{IJ}x^I x^J = g_{ij}y_1^i y_1^j + g_{ij} y_2^i y_2^j + g_{ij} y_3^i y_3^j$, i.e. $G = \begin{pmatrix}
    g & 0 & 0 \\
    0 & g & 0 \\
    0 & 0 & g
\end{pmatrix}$.

The structure constants $c_{ij}^k$ are defined by
\begin{equation}\label{c_ijkdfn}
    [y_i, y_j] = c_{ij}^k y_k \text{ and } c_{ijk} = c_{ij}^{k^\pr} g_{k^\pr k}
\end{equation}

This seems to be the simplest perturbed Gaussian that can be formed from the two tensors $g$ and $c$ which will not vanish due to symmetry consideration. For example, if we merely integrated over one copy, instead of three, of the Lie algebra $\mathfrak{su}(2^n) \cong \R^{4^n - 1}$, using $g_{ij}$ instead of $G_{IJ}$ the integral would become
\begin{equation}
    \tld{F}_k = \int_{\vec{y} \in \R^{4n-1}}\ d\vec{y}\ e^{ik(g_{ij}y^i y^j + c_{ijk}y^i y^j y^k)}.
\end{equation}

But observe that the cubic perturbation $c_{ijk}y^i y^j y^k = 0$ since $c_{ijk} = -c_{jik}$, by the skew symmetry of the Lie bracket $[,]$. The formula in \ref{theintegral} indeed seems the most natural way to build a nontrivial perturbed Gaussian integral from the tensors $g$ and $c$.

Another functional that will be considered is the Euclidean version of the above, where $-i$ in the exponent in (\ref{theintegral}) is replaced by $-1$. 

We should say a little about the symmetries of $F_k$ restricted to the \textbf{kaq} subset $\mcM_{\text{\textbf{kaq}}}^n \subset \mcM^n \coloneqq \{\text{metrics on } \mathfrak{su}(N)\}$. There is the adjoint action of $\operatorname{SU}(N)$ on $\mathfrak{su}(N)$ inducing an action on $\mcM^n$, and since $\mcM^n_\text{\textbf{kaq}}$ is defined via the \emph{existence} of some qubit decomposition, this action preserves $\mcM_\text{\textbf{kaq}}^n$. The isomorphism to $\underbrace{\C^2 \otimes \cdots \otimes \C^2}_{n}$ is merely precomposed by the inverse of the element of $\operatorname{SU}(N)$.  This action, for example has isotopy properly containing $(\underbrace{\operatorname{SU}(2) \times \cdots \times \operatorname{SU}(2)}_{n\text{ copies}})$ on $\mcM^n_{\text{penalty}}$.

For numerical reasons, we are best able to study $f_{k,2}$ for $k$ positive real in the range of $100 \le k \le 1000$ and $f_{k,1}$ for $k$ positive imaginary in a similar range.

Once we locate a locally minimizing metric $g$ we diagonalize it and call the $4^n-1$ eigenvalues the \emph{large spectrum}. Each of the corresponding $4^n-1$ eigenvalues is a $2^n \times 2^n$ skew-Hermitian matrix and we refer to their spectra (which are imaginary) as the \emph{little spectra}. When there is eigenvalue degeneracy, there is additional choice choosing principle axes (eigenvectors of $g_i^j$) and the concomitant little spectra.

There is a very quick and often reliable check that a metric is in $\mcM^n_{\text{\textbf{kaq}}}$: the little spectra should all be consistent with the proposed tensor product structure. This is useful primarily when a non-degenerate eigenvalue in the little spectrum (a 1D eigenspace). A further more definitive test is discussed in section \ref{s4}. In practice we can usually understand the \textbf{kaq} status of the local minima we locate.

The rest of the paper is organized as follows:
\begin{itemize}
    \item \S 2.\ A detailed look at the numerical methods and the machine learning techniques that enabled them.
    \item \S 3. A summary of what computations were carried out and the most significant conclusions for symmetry breaking to \textbf{kaq}s.
    \item \S 4. Summary and outlook.
    \item \S 5. Appendix on the theoretical and linear algebra results.
\end{itemize}

\subsection{The Summary of Results}~
\\
Numerical study of a variety of functionals on inner product on $\mathfrak{su}(4)$ and $\mathfrak{su}(8)$ finds, in addition to an unstable critical point at the bi-invariant Killing form, metric structure often involving multiple local minima. In many cases we have identified these as \textbf{kaq} metrics (see Figures \ref{vensu4} and \ref{vensu8}). Although our study was necessarily limited to very low dimensions, it provides evidence for our proposed SSB scenario. It is furthermore intriguing that the choice of functional does not need to be fine tuned, indicating that perhaps many natural functionals formed from $g_{ij}$ and $c_{ij}^k$ commonly have \textbf{kaq} minima. One could add another layer of probability and consider not just our system Hamiltonian $H$ selected from a metric dependent Gaussian, but also that metric $g_{ij}$ drawn from local minima of functionals $f$ themselves drawn from a distribution $\mcD$ of functionals.

Next, a few disclaimers: Ricci scalar curvature was numerically challenging and we were only able to locate critical points whose locations had previously been determined analytically \cite{jensen1971scalar} and we already knew would be \textbf{kaq}.  These local minima are associated in the two cases with the Lie subalgebras $\mathfrak{so}(4) \subset \mathfrak{su}(4)$ and $\mathfrak{sp}(2) \subset \mathfrak{su}(4)$ respectively. Surprisingly, for the functionals studied on $\mathfrak{su}(4)$, some of the solutions appear numerically to be $\mathfrak{sp}(2)$ Jensen local minima as well. The metric is small on the Lie subalgebra and large on its orthogonal complement. For other local minima for $\mathfrak{su}(4)$ and $\mathfrak{su}(8)$, similar Lie \textbf{sub}algebras have been found that correspond to one (or a combination of some) of the eigenspaces. A more detailed description is given in section \ref{s3}. 

This points to the fact that in addition to the class \textbf{kaq}, there is a class on metrics which we shall call \textbf{sub}$_n$ having the structure described above. Particularly for small $n$, \textbf{sub}$_n$ and \textbf{kaq}$_n$ have considerable intersection among metrics on $\mathfrak{su}(2^n)$. 

We must consider to what extent the evidence  is actually for \textbf{kaq} and not \textbf{sub}. So far, we have not seen local minima in \textbf{sub}$_n  \backslash$ \textbf{kaq}$_n$.
 
In summary, we find considerable numerical evidence to support the plausibility of spontaneous symmetry breaking to \textbf{kaq} metrics, as the prequel to interacting physics.

\section{Numerical Methods}

\subsection{Loss function}
\subsubsection{Gaussian perturbed integral}~
\\
Expanding the perturbative series of (\ref{theintegral}) up to third order, yields a summation of two, four and six vertex trivalent tensor networks, where vertices are labelled by $c$ and edges by $g$ or $g^{-1}$. This expansion follows a similar procedure outlined in \cite[Equation 1.7 onwards]{bar1995perturbative}. We start with the bosonic Gaussian perturbed integral in (\ref{theintegral}), and make a simple change of variable $\vec{x} \to \vec{x} = \sqrt{k}\vec{x}$,
\begin{align}
    F_k = k^{-3(4^n-1)/2}\int_{\mathbb{R}^{3(4^n-1)}} \operatorname{d}\vec{x}\ e^{\frac{i}{2}G_{IJ}x^Ix^J} e^{\frac{i}{\sqrt{k}}c_{ijk} y_1^iy_2^jy_3^k } = 
    \\ k^{-3(4^n-1)/2}\int_{\mathbb{R}^{3(4^n-1)}} \operatorname{d}\vec{x} \ e^{\frac{i}{2}G_{IJ}x^Ix^J} \sum_{m=0}^{\infty } \frac{i^m}{m!k^{m/2}} (c_{ijk}y_1^iy_2^jy_3^k)^m
\end{align}
Focusing on the $m$-th term of the expansion gives a Gaussian integral
\begin{align}
    \int_{\mathbb{R}^{3(4^n-1)}} \operatorname{d}\vec{x} \ e^{\frac{i}{2}G_{IJ}x^Ix^J} (c_{ijk}y_1^iy_2^jy_3^k)^m 
\end{align}
which can be computed using standard methods
\begin{align}
    [(c_{ijk} \frac{-i\partial}{\partial V_1^i}\frac{-i\partial}{\partial V_2^j}\frac{-i\partial}{\partial V_3^k})^m \int_{\mathbb{R}^{3(4^n-1)}} \operatorname{d}\vec{x} \ e^{\frac{i}{2}(g_{ij}y_1^i y_1^j + g_{ij} y_2^i y_2^j + g_{ij} y_3^i y_3^j) + V_1^iy_1^i+V_2^jy_2^j + V_3^ky_3^k}]_{\vec{V} = 0} \\
    \propto [(c_{ijk} \frac{-i\partial}{\partial V_1^i}\frac{-i\partial}{\partial V_2^j}\frac{-i\partial}{\partial V_3^k})^m e^{-\frac{i}{2}G^{IJ}V_IV_J }]_{\vec{V}=0}
\end{align}
Writing $e^{-\frac{i}{2}G^{IJ}V_IV_J}$ as a Taylor series $\sum_p \frac{(-\frac{i}{2}G^{IJ}V_IV_J)^p}{p!}$, since the exponent $G^{IJ}V_IV_J$ is a quadratic expression, the only nonzero terms after differentiation and setting $\vec{V}=0$ correspond to the power $p = \frac{3m}{2}$, i.e.
\begin{align}
    (c_{ijk} \frac{-i\partial}{\partial V_1^i}\frac{-i\partial}{\partial V_2^j}\frac{-i\partial}{\partial V_3^k})^m (-\frac{i}{2}G^{IJ}V_IV_J)^{\frac{3m}{2}}.
\end{align}
Indeed all other terms are zero, when $p<\frac{3m}{2}$, as there are $3m$ partials and less ($<2\times \frac{3m}{2}$) variables, and when $p>\frac{3m}{2}$, as $\vec{V}=0$.

By following the power of imaginary number $i$ in the previous equations, we observe that $\operatorname{Im}(F_k) = f_{k,2}$ corresponds to $m \equiv 2 \pmod 4$, while the rest is $\operatorname{Re}(F_k) = f_{k,1}$.  Similarly, the Euclidean version of this integral has alternating signs $\pm1$ depending on $m \pmod 4$.

Our numerical experiments are based on $m=2,6$ and on $m = 2,4$ for the Euclidean version. 

We can show that the expression above can be viewed as a contraction of a trivalent network without any loops, and $m$ vertices $c_{ijk}$ and edges $g^{ii},g^{jj},g^{kk}$. Note each partial needs to be paired with its corresponding term in $(-\frac{i}{2} G^{IJ}V_IV_J)^{\frac{3m}{2}}$. Let us fix a $j$ and consider the example of $\frac{-i\partial}{\partial V_2^j}$ which is associated with some term $c_{-j-}$. Inside $(-\frac{i}{2} G^{IJ}V_IV_J)^{\frac{3m}{2}}$, the term $V_2^j$ comes with another $g^{jj'}V_2^{j'}$. Therefore, there can not be any loops, or any pairing between two different indices, like $i$ and $j$. As $V_2^{j'}$ is similarly paired with $\frac{-i\partial}{\partial V_2^{j'}}$, which is associated with some term $c_{-j'-}$, we get an \textit{edge} connecting the two \textit{vertices} $c_{-j-}$ and $c_{-j'-}$ with $g^{jj'}$ labeling that edge. 

We note a further simplification, by using (\ref{c_ijkdfn}), allowing us to conclude that vertices can be labelled by $c_{ij}^k$ instead of $c_{ijk}$, while edges are labelled by $g^{ii'},g^{jj'}$ and $g_{kk'}$ instead of $g^{kk'}$. The reason for the change in the latter edge type is that when two vertices $c_{ijk}$ and $c_{i'j'k'}$ are connected along the $kk'$ indices, two factors of $g_{kk'}$ come from $c_{ijk}$ and $c_{i'j'k'}$ (thus transforming them to $c_{ij}^k,c_{i'j'}^{k'}$), and one gets cancelled by $g^{kk'}$ coming from $g^{kk'}V_3^kV_3^{k'}$.

Each diagram also comes with a coefficient. The coefficient can be computed by following the scalars in equations above for each $m$, and considering all permutations to each diagram labelling.

\begin{figure}[h]
    \centering
\begin{tikzpicture}
\Vertex[x=1,label=$c$]{A}
\Vertex[x=1,y=-2,label=$c$]{B}
\Edge[label=$k$, color = red](A)(B)
\Edge[bend=65,label=$i$](A)(B)
\Edge[bend=-65,label =$j$](A)(B)
\end{tikzpicture}
    \caption{Theta diagram. All diagrams are trivalent networks without any loop, and vertices are the structure constants $c_{ij}^k$. Each vertex has indices $i,j,k$ which are paired with their counterpart in another vertex. This pairing is done using $g$ along edge of type $k$ (colored red) and $g^{-1}$ for type $i$ and $j$.}
    \label{Thetadiagram}
\end{figure}
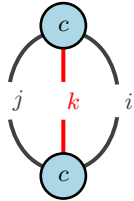

\begin{figure}[h]
    \centering
\begin{tikzpicture}
\begin{scope}[scale=1.5,shift={(-5,0)}]
\Vertex[label=$c$]{A}
\Vertex[x=1,y=0.8,label=$c$]{B}
\Vertex[x=0.5,y=2,label=$c$]{C}
\Vertex[x=2,label=$c$]{D}
\Vertex[x=3,y=0.8,label=$c$]{E}
\Vertex[x=2.5,y=2,label=$c$]{F}
\Edge[label=$k$, color =red](A)(B)
\Edge[label=$j$](B)(C)
\Edge[label=$i$](C)(A)
\Edge[label=$k$, color =red](D)(E)
\Edge[label=$j$](E)(F)
\Edge[label=$i$](F)(D)
\Edge[label=$j$](A)(D)
\Edge[label=$i$](B)(E)
\Edge[label=$k$, color =red](C)(F)
\end{scope}
\begin{scope}[scale=1.5]
\Vertex[label=$c$]{A}
\Vertex[x=2,label=$c$]{B}
\Vertex[x=4,label=$c$]{C}
\Vertex[y=2,label=$c$]{D}
\Vertex[x=2,y=2,label=$c$]{E}
\Vertex[x=4,y=2,label=$c$]{F}
\Edge[label=$k$, color =red,distance=.2](A)(D)
\Edge[label=$j$,distance=.2](A)(E)
\Edge[label=$i$,distance=.2](A)(F)
\Edge[label=$i$,distance=.2](B)(D)
\Edge[label=$k$, color =red,distance=.2](B)(E)
\Edge[label=$j$,distance=.2](B)(F)
\Edge[label=$j$,distance=.2](C)(D)
\Edge[label=$i$,distance=.2](C)(E)
\Edge[label=$k$, color =red,distance=.2](C)(F)
\end{scope}
\begin{scope}[scale=1.5,shift={(-5,-3)}]
\Vertex[label=$c$]{A}
\Vertex[x=2,label=$c$]{B}
\Vertex[y=2,label=$c$]{C}
\Vertex[y=1,label=$c$]{E}
\Vertex[x=2,y=2,label=$c$]{D}
\Vertex[x=2,y=1,label=$c$]{F}
\Edge[label=$k$, color =red](A)(E)
\Edge[label=$k$, color =red](B)(F)
\Edge[label = $j$](C)(E)
\Edge[label = $j$](D)(F)
\Edge[bend=45,label=$i$](A)(B)
\Edge[bend=-45,label=$j$](A)(B)
\Edge[bend=45,label=$i$](C)(D)
\Edge[bend=-45,label=$k$, color =red](C)(D)
\Edge[label=$i$](E)(F)
\end{scope}
\begin{scope}[scale=1.5,shift={(1,-3)}]
\Vertex[label=$c$]{A}
\Vertex[x=1.5,label=$c$]{B}
\Vertex[y=2,label=$c$]{C}
\Vertex[x=-0.5,y=1,label=$c$]{E}
\Vertex[x=1.5,y=2,label=$c$]{D}
\Vertex[x=2,y=1,label=$c$]{F}
\Edge[bend = 45, label=$k$, color =red](A)(E)
\Edge[bend =45, label = $i$](B)(F)
\Edge[label = $j$](C)(E)
\Edge[label = $j$](D)(F)
\Edge[bend=45,label=$k$, color =red](F)(B)
\Edge[label=$j$](A)(B)
\Edge[bend=45,label=$k$, color =red](C)(D)
\Edge[bend=-45,label=$i$](C)(D)
\Edge[bend= 45, label=$i$](E)(A)
\end{scope}
\begin{scope}[scale=1.5,shift={(-2.5,-6)}]
\Vertex[label=$c$]{A}
\Vertex[x=1,y=0.8,label=$c$]{B}
\Vertex[x=0.5,y=2,label=$c$]{C}
\Vertex[x=2,label=$c$]{D}
\Vertex[x=3,y=0.8,label=$c$]{E}
\Vertex[x=2.5,y=2,label=$c$]{F}
\Edge[label=$k$, color =red](A)(B)
\Edge[label=$j$](B)(C)
\Edge[label=$i$](C)(A)
\Edge[label=$k$, color =red](D)(E)
\Edge[bend = 45, label=$j$](E)(F)
\Edge[bend = 45, label=$i$](F)(E)
\Edge[label=$j$](A)(D)
\Edge[label=$i$](B)(D)
\Edge[label=$k$, color =red](C)(F)
\end{scope}
\end{tikzpicture}
    \caption{From left to right, top to bottom: Prism, $K_{3,3}$, Extended tincan, Necklace, and TetraTheta diagrams, each with some sample labeling. Red lines are labelled by $g$ and black lines by $g^{-1}$.}
    \label{6vertexdiagrams}
\end{figure}
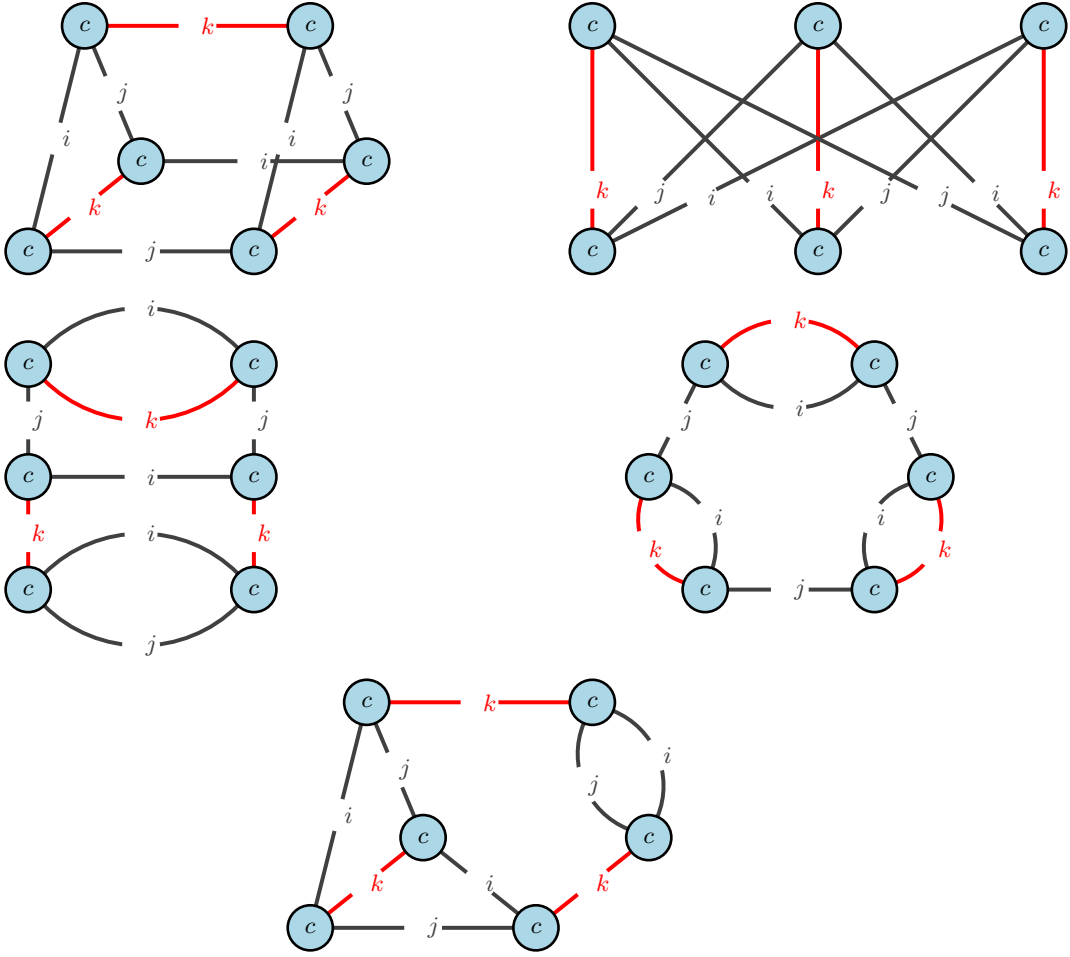

We list all diagrams for $m=2,4,6$ in Figures \ref{Thetadiagram}, \ref{TincanTetrahedrondiagram}, \ref{6vertexdiagrams} along with their coefficients.

For $m=2$, there is only the diagram
\begin{itemize}
    \item Theta
\end{itemize}
with coefficient $\frac{1}{2!}$.
For $m=4$, there is a common coefficient $\frac{1}{4!}$.
\begin{itemize}
    \item Two thetas
    \item Tincan
    \item Tetrahedron
\end{itemize}
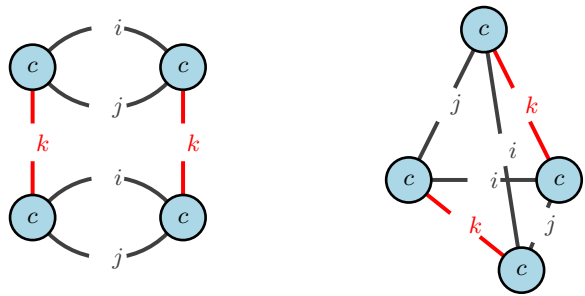
\begin{figure}[h]
    \centering
\begin{tikzpicture}
\begin{scope}[shift={(-5,-0.5)}]
\Vertex[label=$c$]{A}
\Vertex[x=2,label=$c$]{B}
\Vertex[y=2,label=$c$]{C}
\Vertex[x=2,y=2,label=$c$]{D}
\Edge[label=$k$, color =red](A)(C)
\Edge[label=$k$, color =red](B)(D)
\Edge[bend=45,label=$i$](A)(B)
\Edge[bend=-45,label=$j$](A)(B)
\Edge[bend=45,label=$i$](C)(D)
\Edge[bend=-45,label=$j$](C)(D)
\end{scope}
\Vertex[label=$c$]{A}
\Vertex[x=2,label=$c$]{B}
\Vertex[x=1.5, y=-1.2,label=$c$]{C}
\Vertex[x=1,y=2,label=$c$]{D}
\Edge[label=$k$, color =red](A)(C)
\Edge[label=$k$, color =red](B)(D)
\Edge[label=$i$](A)(B)
\Edge[label=$j$](A)(D)
\Edge[label=$i$](C)(D)
\Edge[label=$j$](B)(C)
\end{tikzpicture}
    \caption{Tincan and Tetrahedron with some sample labeling. Red lines are labelled by $g$ and black lines by $g^{-1}$.}
    \label{TincanTetrahedrondiagram}
\end{figure}
For $m=6$, there is a common coefficient $\frac{1}{6!}$.
\begin{itemize}
    \item Three thetas
    \item Tincan and theta
    \item Tetrahedron and theta
    \item Prism
    \item $K_{3,3}$
    \item Extended tincan
    \item Necklace
    \item TetraTheta
\end{itemize}

Hence, the functions of study are 
\begin{align}\label{F_26}
    F_{26}(c,g,k) = \frac{1}{6!} (\text{sum of 6-vertex diagrams}) - \frac{k^2}{2!}(\text{the 2-vertex diagram Theta})
    \\\label{F_24}
    F_{24}(c,g,k) = \frac{1}{4!}(\text{sum of 4-vertex diagrams}) - \frac{k}{2!} (\text{the 2-vertex diagram Theta})
\end{align}
where the power of $k$ for each $m$ is $-\frac{3(4^n-1)+m}{2}$, thus the ratio yields the term $k^2$ in $F_{26}$ and $k$ in $F_{24}$ for $m=2$.

\begin{rmk}
As mentioned in the introduction, and illustrated in experiments and in Appendix \ref{tensorsconvex}, the tensor contractions above are each convex with minimum at Id. This is why it is expected that a signed sum of these diagrams gives local minima around a local maximum at Id (Figure \ref{mexicanhat}).
\end{rmk}
In order to fix the volume by restricting to the space of metrics with $\det(g) = 1$, we add the term $(\det(g)-1)^2$ with a high enough coefficient to the loss function. The more obvious normalization is a suitable fractional power of $\frac{1}{\det(g)}$. However, not only this power is different for networks with different number of vertices, but also, such an expression leads to more severe numerical round-off errors because of the fractional power, as well as a more involved formula for the gradient than $(\det(g)-1)^2$ (see Remark \ref{whyginsteadofginverse}).

Furthermore, in many instances, the value of $F_{26},F_{24}$ is too high, which makes the use of scaling factors for the functional and $\det(g)$ necessary in order to avoid numerical instability, sometimes leading to $\det(g)$ being sent to zero.

The general form of the loss function is:
\begin{align}\label{L_26}
    L_{26}(c,g,k) = r_1^{-1}F_{26}(c,g,k) + r_2(\det(g)-1)^2, \\\label{L_24}
    L_{24}(c,g,k) = r_1^{-1}F_{24}(c,g,k) + r_2(\det(g)-1)^2,
\end{align}
where $r_1 \ge 1, r_2 >>1$. The term \textbf{solution} will be exclusively used to refer to the critical points of the loss function found through gradient descent, which experimentally (through many random small perturbations) have been checked to be local minima. This last check is necessary, as we are looking for local minima of $F_{26},F_{24}$ inside the $\det(g)=1$ subspace, and it is not immediately clear that the set of such local minima is equal to that of $L_{26},L_{24}$ in the space of all metrics $g$.~
\\
\subsubsection{Ricci scalar curvature}~
\\
The Ricci scalar curvature for all unimodular Lie algebras can be computed as follows \cite{jensen1971scalar}:
\begin{align}\label{scalarcurvature}
    R = -\frac{1}{2} \sum_{i,i',j,k}c_{ij}^kc_{i'k}^j g_{ii'} -\frac{1}{4}\sum_{i,j,k}\sum_{i',j',k'} c_{ij}^kc_{i'j'}^{k'}g_{ii'}g_{jj'}g^{kk'}
\end{align}
which in diagrammatic notation Figure \ref{Rscalarcurvature}.
\begin{figure}[h]
    \centering
\begin{tikzpicture}
\Text[x=-2,y=-1]{$R=-\frac{1}{2}$}
\Vertex[label=$c$]{A}
\Vertex[y=-2,label=$c$]{B}
\Vertex[shape=coordinate, y=-1, label = $\operatorname{Id}_{{\color{red}k}j}$,color = white]{midAB}
\Vertex[shape=coordinate, y=-1.25]{midABlow}
\Vertex[shape=coordinate, y=-0.75]{midABhigh}
\Edge[color=red](midABlow)(B)
\Edge(midABhigh)(A)
\Edge[bend=65,label=$g_{ii'}$](A)(B)
\Vertex[shape=coordinate, y=-1, x= -0.7575, label = $\operatorname{Id}_{j\color{red}k}$,color = white]{midABleft}
\Vertex[shape=coordinate, y=-1.25,x= -0.7575]{midABleftlow}
\Vertex[shape=coordinate, y=-0.75,x= -0.7575]{midABlefthigh}
\Edge[bend=32.5](B)(midABleftlow)
\Edge[bend=-32.5,color=red](A)(midABlefthigh)
\begin{scope}[shift={(3,0)}]
\Text[x=-1.5,y=-1]{$-\frac{1}{4}$}
\Vertex[label=$c$]{A}
\Vertex[y=-2,label=$c$]{B}
\Edge[label = $g^{kk'}$, color = red](A)(B)
\Edge[bend=65,label=$g_{ii'}$](A)(B)
\Edge[bend=-65,label=$g_{jj'}$](A)(B)
\end{scope}
\end{tikzpicture}
    \caption{Ricci scalar curvature.}
    \label{Rscalarcurvature}
\end{figure}

For $\mathfrak{su}(2^n)$, this can be simplified to
\begin{align}
    R = \frac{1}{2} \sum_{i,j,k}(c_{ij}^{k})^2 g_{ii} -\frac{1}{4}\sum_{i,j,k}\sum_{i',j',k'} c_{ij}^kc_{i'j'}^{k'}g_{ii'}g_{jj'}g^{kk'}.
\end{align}
To find the critical points, we need to minimize $||\nabla R||$ which can be computed explicitly using the formula above. Minimizing $||\nabla R||$ through gradient descent is too slow, likely because the functional has a shape very close to flat. In this kind of situation, in machine learning applications, one method that generally provides better results is the \textit{evolutionary} method. The idea is to do an approximate gradient descent, but much faster and with more potential in finding lower minima in almost flat-shaped graphs.

Instead of computing a gradient, we perturb by small amount the given point (\textit{parent}) and compute the value of $M$ many of its perturbations (\textit{children}). The child with the smallest evaluation is chosen as the new parent. In our application, we choose $M=20$ and perturb each entry of the parameter by using a uniform distribution on $(1,1/30)$.~
\\
\subsection{Initialization of the gradient descent}~
\\
Different initializations of the metric $g$ lead to the discovery of different local minima of the loss function. In addition, some forms of initializations are preserved under gradient descent. We need to discuss some terminology and lemmas to discuss the initializations.

\begin{dfn}
Given a metric $g$, the \textit{degeneracy pattern} is the set $\{d_1,\ldots,d_t\}$ of the dimensions of its eigenspaces. An \textbf{ordered} degeneracy pattern is a tuple $(d_1,\ldots,d_t)$ where the corresponding eigenvalues $0<\lambda_{d_1} < \ldots < \lambda_{d_t}$ are in ascending order.
\end{dfn}

As an example, the ordered \textbf{penalty} degeneracy pattern (or just ordered penalty pattern) is $(6,9),(9,27,27)$ for $\mathfrak{su}(4),\mathfrak{su}(8)$, respectively.

The gradient flow preserves particular degeneracy patterns related to the \textbf{Clifford group}: Given $n$ qubits, the Clifford group $C_n=\{V \in U_{2^n} | V \textbf{P}_n V^\dagger = \textbf{P}_n \}$ is the group of unitary gates that normalize the Pauli group $\textbf{P}_n = \{e^{i\theta\pi/2} \sigma_{i_1} \otimes \cdots \otimes\sigma_{i_n} | \theta \in \{0,1,2,3\}, i_j \in \{0,x,y,z\}\}$, where $\sigma_0 = \operatorname{Id}$. 

Let a subgroup $H<C_n$ act on $\textbf{P}_n/\{\pm1, \pm i\}$, where we identify words that are proportional to each other. Consider the orbits of $H$, which also provide a decomposition of the Pauli word basis $\textbf{PB}_n$ of $\mathfrak{su}(2^n)$, thus determining a degeneracy pattern. We have the following theorem.
\begin{thm}[Appendix \ref{app_preserv_diag_thm}] \label{preservedegeneracypattern}
Let $D$ be diagonal with a degeneracy pattern determined according to the orbits of a subgroup $H<C_n$ as explained above. The gradient flow preserves not only the diagonal form of the metric, but also the degeneracy pattern modulo the merging of the eigenspaces, e.g $\{d_1,d_2,d_3\}$ could become $\{d_1,d_2+d_3\}$.
\end{thm}

\begin{rmk}
As a direct corollary, gradient descent reaches a critical point (experimentally checked to be local minima) with that degeneracy pattern modulo merging.
\end{rmk}

We can now discuss different types of initialization to start the gradient descent with. All such initialization are normalized to have $\det =1$, and determine the set of parameters on which gradient descent is performed.

\begin{itemize}
    \item \textbf{DiagPerturbId}: A uniform random diagonal perturbation of the local maxima at identity, the set of parameters being the diagonal entries.
    \item \textbf{PenMetricOmega}: Penalty metric with gradient on the weight parameter $w=e^r$ in (\ref{penalty_metric}), which means the local minima are found in the space of penalty metrics. Thus, a solution may not necessarily be a local minimum in the space of all metrics (but see Remark \ref{patternequalsomega}). This is to search for the minimum that the penalty metric can achieve and compare it to the lowest local minima found in other ways.
    \item \textbf{PenMetricPattern} : Penalty metric with gradient descent on $w_1,...,w_n $ where $w_i$ is the entry for Pauli word of weight $i$. All $w_i$ are initialized using a random $w$, i.e. $w_i = w^i$ at the start. Once the local minimum is reached, we can observe how far the local minima found is from a penalty metric by comparing the ratios $w_{i+1}/w_i$.
    \item \textbf{GenPerturbId}: A general (nondiagonal) random Gaussian perturbation of identity, the set of parameters being entries of the matrix.
    \item \textbf{GenMetric}: A random metric, the parameters being entries initialized by standard normal distribution.
\end{itemize}

\begin{rmk}\label{patternequalsomega}
For $\mathfrak{su}(4)$, \textbf{PenMetricOmega} is effectively the same as the more general \textbf{PenMetricPattern}: There is only one ratio  $(w_2/w_1)$ to calculate and it gives the weight parameter $w$. On the other hand, a penalty pattern for $\mathfrak{su}(8)$ could experience a merging of eigenspaces or lead to a solution where the eigenvalues do not represent a penalty pattern (i.e. $w_2/w_1 \neq w_3/w_2$), and all local minima found with this initialization for $\mathfrak{su}(8)$ are as such.
\end{rmk}~
\\
\subsection{Adam optimization algorithm}~
\\
The Adam optimization algorithm \cite{kingma2014adam} has become one of the most used algorithms for deep learning and more generally machine learning applications. Similar to the vanilla stochastic gradient descent, it is used to update parameters of a functional iteratively. However, it has many advantages compared to the stochastic gradient descent, and has been experimentally shown to lead to lower minima faster compared to other stochastic optimization algorithms \cite{kingma2014adam}. Most important to our application, it satisfies some desirable characteristics:
\begin{itemize}
    \item Computationally efficient with little memory requirements.
    \item Invariant to diagonal rescale of the gradients.
    \item Appropriate for problems with sparse gradients.
    \item Hyper-parameters have intuitive interpretation and typically require little tuning.
\end{itemize}

In contrast to stochastic gradient descent, Adam has a learning rate for each parameter which is separately adapted during learning. The name Adam comes from \textit{adaptive moment estimation}, as it computes individual adaptive learning rates for each parameter from estimates of first and second moments of the gradients. This helps with sparse gradients situations and cases where the gradient is highly noisy (like in nonstationary settings). For our purpose, we would like an algorithm that avoids saddle points and has a better chance of finding the more interesting local minima by not getting stuck in small basins.

\begin{figure}[h]
    \centering
    \includegraphics[width=0.9\textwidth]{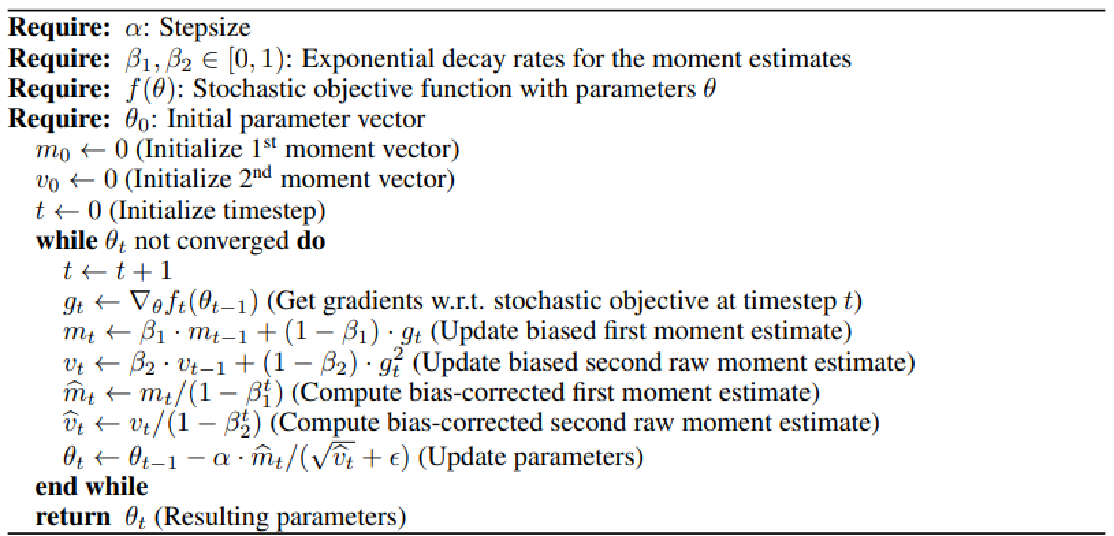}
    \caption{Adam algorithm as in \cite{kingma2014adam}}
    \label{Adam_algo}
\end{figure}

Figure \ref{Adam_algo} shows the implementation of the algorithm.
The hyperparameters of Adam with their default values are described below.
\begin{itemize}
    \item $\alpha$ (default $0.001$): The learning rate (LR) or stepsize. The proportion that weights are updated. We prefer small values to have a more thorough search and find more local minima.
    \item $\beta_1$ (default $0.9$): The exponential decay rate for the first moment estimates. This hyperparameter like the next ones required no tuning.
    \item $\beta_2$ (default $0.999$): The exponential decay rate for the second-moment estimates. In regions with a small or sparse gradients, $\beta_2$ should be close to $1$.
    \item $\epsilon$ (default $10^{-8}$): This is to avoid any division by zero in the implementation.
\end{itemize}~
\\
\subsubsection{Scheduling and hyperparameters}~
\\
LR decay can be used with Adam. In all cases, LR is $10^{-3}$ but is halved during the gradient descent every 20000 steps, as the algorithm struggles to converge in its final stages, as it jumps over and back on lower minima basins due to larger than appropriate LR. All other hyperparameters are taken with their default values. Scaling factors mostly depend only on the type of the integral and value of $k$. The scaling factors chosen for each initialization are given in Tables \ref{scalingfactorstable26} and \ref{scalingfactorstable24}. Generally, we prefer to choose higher $r_1$ to scale $F_{26}$ or $F_{24}$ if they attain very high values, which happens as $k$ grows larger. If they attain high values but we can still be confident about the accuracy of the computations, $r_1$ is set to $1$ while $r_2$ is chosen high enough ($10^6$) to avoid $\det(g)$ vanishing. 

\begin{table}[h]
    \centering
    \begin{center}
    \begin{tabular}{ |c|c|c|c|c| } 
    \hline
    $k=100,200$ & DiagPerturbId & GenPerturbId &  PenMetricPattern \\
    \hline
    $r_1$ & $1$ & $1$ & $1$\\ 
    $r_2$ & $10^6$ & $10^6$ & $10^6$\\
    \hline
    \end{tabular}
    \begin{tabular}{ |c|c|c|c|c| } 
    \hline
    $k=500,1000$ & DiagPerturbId & GenPerturbId & PenMetricOmega & PenMetricPattern \\
    \hline
    $r_1$ & $10^4$ & $10^4$ & $10^4$ & $10^4$\\ 
    $r_2$ & $10^3$ & $10^4$ & $10^3$ & $10^3$ or $10^4$\\
    \hline
    \end{tabular}
    \end{center}
    \caption{The scaling factors for $L_{2,6}$ for $\mathfrak{su}(4) (k=100,200), \mathfrak{su}(8) (k=500,1000)$.}
    \label{scalingfactorstable26}
\end{table}

\begin{table}[h]
    \centering
    \begin{center}
    \begin{tabular}{ |c|c|c|c|c| } 
    \hline
    $k=100,200$ & DiagPerturbId & GenPerturbId & PenMetricPattern \\
    \hline
    $r_1$ & $10$ & $10$ & $10$\\ 
    $r_2$ & $10^3$ & $10^3$ & $10^3$\\
    \hline
    \end{tabular}
    \begin{tabular}{ |c|c|c|c| } 
    \hline
    $k=1000$ & DiagPerturbId & GenPerturbId &  PenMetricPattern \\
    \hline
    $r_1$ & $10$ & $10$ & $10$\\ 
    $r_2$ & $10^3$ & $10^3$ &  $10^3$\\
    \hline
    \end{tabular}
    \end{center}
    \caption{The scaling factors for $L_{2,4}$ for $\mathfrak{su}(4) (k=100,200), \mathfrak{su}(8) (k=1000)$.}
    \label{scalingfactorstable24}
\end{table}

\section{Optimization Results and Conclusions}\label{s3}

We shall make a few remarks before discussing the results.
\begin{rmk}
In tables above, \textbf{GenMetric} is absent. It was observed through experiments that they 
\begin{itemize}
    \item did not give any new local minima (in terms of degeneracy pattern), or
    \item they did not converge, as either the values of the functionals were generally too high for us to have confidence in the numerics, or the complicated scheduling necessary in the hyperparameter tuning made it too hard to have a stable gradient descent.
\end{itemize} 
\end{rmk}
\begin{rmk}
As mentioned in Remark \ref{patternequalsomega}, for $\mathfrak{su}(4)$, ``\textbf{PenMetricPattern} = \textbf{PenMetricOmega}'', but for $\mathfrak{su}(8)$, they are gradient descents in different spaces. In the case of $L_{2,4}$ (Table \ref{scalingfactorstable24}), we could not collect any nontrivial result for $\mathfrak{su}(8)$ by using \textbf{PenMetricOmega}, hence why its column is absent.
\end{rmk}
\begin{rmk}
Our experiments involved two values of $k$ although the values in-between were tried to some extent, and did not deliver any new results. If a degeneracy pattern reappears for higher values of $k$, it is always the case that eigenvalues are further apart.
\end{rmk}
We assess our solutions using the following type of evidences, listed in the order used in the paper.
\begin{itemize}
    \item Finding Lie subalgebra structure by computing the brackets of (a collection of) eigenspaces. 
    \item We propose two different ways for providing evidence/proof regarding the existence of a map $j$ as defined in \ref{dfnkaq}. In this paper, we explore the first one to some extent:
    \begin{enumerate}
    \item An evidence for the existence of $j$ can be found by looking for what is called \textit{pattern breaking}. First one finds a partition of the given degeneracy pattern that has the same dimensions as the penalty pattern; e.g. for $\mathfrak{su}(4)$, the pattern $(3,1,1,8,2)$ (see further below) has one partition such as $(\{3,1,2\}, \{1,8\})$ where the dimensions in each set add up to the penalty pattern $(6,9)$. Next, we check if the Lie algebra generated by the set of eigenspaces ($\{3,1,2\}$ in the example) corresponding to the weight one Pauli words is isomorphic to the  Lie subalgebra $\mathfrak{su}(2)\oplus \ldots \oplus \mathfrak{su}(2)$ ($n=2$ times in the example). This ensures that there exists a map $j$, at least for the restriction to weight one Pauli words. The final step, not carried out in this work, would be to see if the rest of the eigenvectors can also be mapped under such a $j$ to a local tensor product form.
    \item A more direct approach is a gradient descent in the space of all Lie algebra isomorphisms $j^{-1}$. We can check whether $j^{-1}(H_i)$ is in tensor product form by computing the entanglement entropy and aim to minimize the overall entanglement entropy of $j^{-1}(H_i)$s. 
    
    Consider the case of $\mathfrak{su}(4)$ where we would like to measure the distance of $j^{-1}(H_i)$ to a decomposition into a tensor product $O_1 \otimes O_2$. Fixing $j$ (or $J$), determines an isomorphism of the underlying $\mathbb{C}^4$ (on which $\mathfrak{su}(4)$ are the skew-Hermitian transformations) into $\mathbb{C}^4 \cong \mathbb{C}_1^2 \otimes \mathbb{C}_2^2$. Think of $j^{-1}(H_i)$ as a vector in $\mathbb{C}_1^2 \otimes \mathbb{C}_2^2 \otimes (\mathbb{C}_1^2)^\star \otimes (\mathbb{C}_2^2)^\star$ so that the partial trace $\rho_i = \operatorname{Tr}_{\mathbb{C}_1^2}(j^{-1}(H_i))\in \mathbb{C}_2^2 \otimes (\mathbb{C}_2^2)^\star$ under the natural map. 
    
    By computing the entanglement entropy of $j^{-1}(H_i)$ which is $S(j^{-1}(H_i)):= S_i = \operatorname{Tr}(\rho_i \log \rho_i)$, we can define the loss function as $\sum_{i=1}^{\dim \mathfrak{su}(4) = 15} S_i$ and aim to minimize it over $\{j\}$. If $\sum S_i$ can be made ``close'' to zero, then we conclude that $\{H_i\}_{i=1}^{\dim \mathfrak{su}(4) = 15}$, is \textbf{kaq} up to some error (depending on how ``close'' to zero); i.e. the eigenvectors for $g$ have attained something close to a tensor product form. Note that for higher number of qubits, like in $\mathfrak{su}(8)$, this test needs to be performed successively to prove decomposition into $O_1 \otimes O_2\otimes O_3$.
    \end{enumerate}
    \item If we find a nondegenerate eigenvalue, we make a little spectra test on the corresponding skew-Hermitian matrix, to find if the spectrum can rule out that the matrix is not in the form of a tensor product.
    \item In the case of \textbf{PenMetricPattern}, we make a direct inspection and look at the ratios of eigenvalues of different weight spaces (Remark \ref{patternequalsomega}).
    \item In the case of \textbf{PenMetricOmega}, we check if the solution is a local minima in the space of all metrics, and if the weight parameter is smaller than one.
\end{itemize}
We used the \href{https://github.com/google/TensorNetwork}{\textbf{TensorNetwork}} package to compute the diagrams and \href{https://github.com/pytorch/pytorch}{\textbf{PyTorch}} to do gradient descent using Adam. For the hardware, we used one NVIDIA Tesla V100 GPU. Most simulations took at most 2 hours to converge, with the diagonal initializations converging faster.

\begin{rmk}\label{whyginsteadofginverse}
PyTorch uses Automatic differentiation (Autograd) to compute gradient of a function $f$. It does so by first building a computational tree of $f$ using simpler functions, such as linear, trigonometric, etc., for which the gradient formula is built into the program and can be calculated exactly. Then, it computes the gradient using the chain rule.

As such, to make gradient descent computations faster on the computer, \textbf{we replace $g$ with $g^{-1}$}, making sure that one type of edge ($k$) gets labelled by $g^{-1}$, and two labelled by $g$. \textit{Thus, when we examine the results (eigenvalues of $g$), we will need to take the inverse again}. Note that as the parameters of our program are the entries of $g$, the computation for gradients of terms involving $g$ is far simpler than those involving $g^{-1}$, as the formula for the latter involves many long expressions in terms of $g$ entries (determinants of minors and division by $\det(g)$). Therefore, it makes sense to make such a replacement as it allows us to cut the number of edges labelled by $g^{-1}$ in half (and double the number of those labelled by $g$).
\end{rmk}~
\\
\subsection{\textbf{DiagPerturbId} and \textbf{GenPerturbId} results}
\begin{figure}
    \centering
    \includegraphics{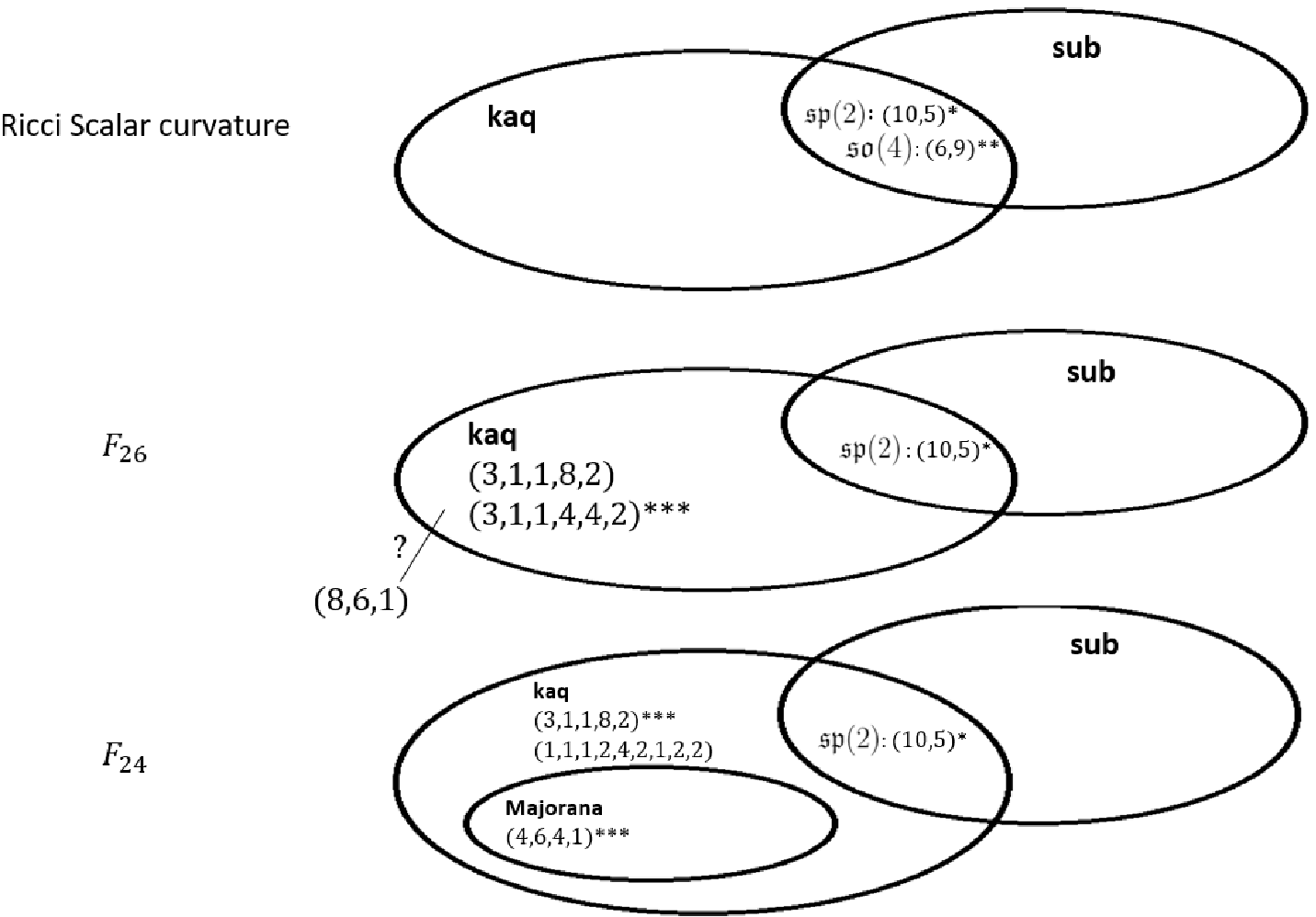}
    \caption{Patterns found for the $\mathfrak{su}(4)$ for each functional and their possible classification. *: All $\mathfrak{sp}(2)$s are found in diagonal solutions. **: The penalty metric. ***: The ordered patterns are slightly modified given that the eigenvalues were numerically close; this is done to represent the ordered degeneracy pattern in a way that relates it to a previously found ($(3,1,1,8,2)$) or well-known (Majorana) pattern. ?: The classification of the pattern $(8,6,1)$ is unclear.}
    \label{vensu4}
\end{figure}

\begin{figure}
    \centering
    \includegraphics{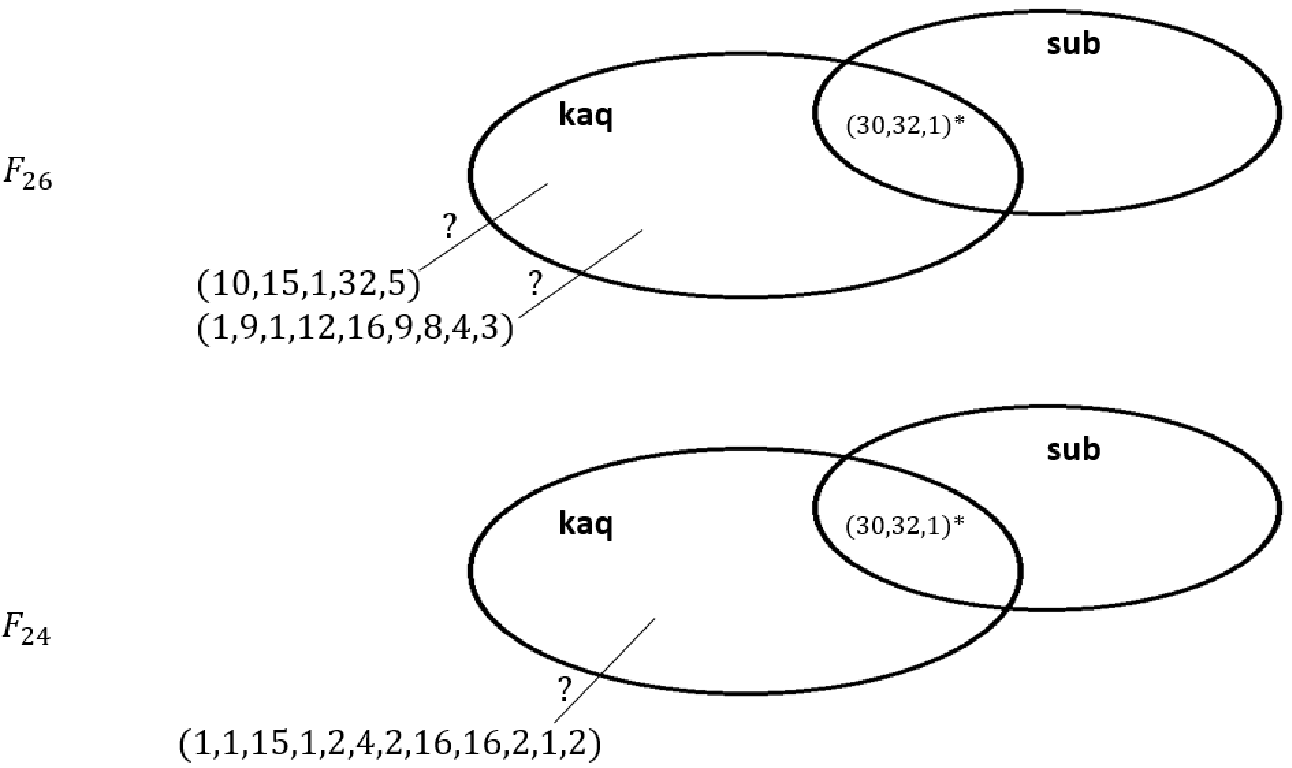}
    \caption{Patterns found for the $\mathfrak{su}(8)$ for each functional. *: The diagonal patterns lead to very similar $(30,32,1)$ pattern in each functional (see the text for details). ?: The classification of all the other patterns such as $(10,15,1,32,5)$ is unclear.}
    \label{vensu8}
\end{figure}

~
\\
There are different degeneracy patterns among the solutions of \textbf{DiagPerturbId} and \textbf{GenPerturbId}. We will list the (lowest) local minima ordered degeneracy patterns along with eigenvalues for the highest $k$ (if it appears for both $k$), and provide explanation regarding the \textbf{kaq}/\textbf{sub}-ness of the metrics. In these lists, LM = local minima and LLM = lowest local minima. A summary of the results is presented in Figures \ref{vensu4} and \ref{vensu8}.

1- $L_{2,6}$ and $\mathfrak{su}(4)$ with $k = 100,200$:
\begin{itemize}
    \item LM: The second pattern occurs exclusively for $k=100$ while the last two patterns do so for $k=200$.
    \begin{enumerate}
        \item $(10,5), (0.315, 10.079)$
        \item $(3,1,1,8,2), (0.650, 0.678, 0.686, 0.951, 3.417)$
        \item $(1,3,1,4,4,2),(0.309, 0.311, 0.333, 0.381, 3.424, 10.526)$
        \item $(8,6,1),(0.396,2.523,44.247)$
    \end{enumerate} 
    \item LLM for both $k$: $(10,5)$ where the 10D eigenspace is $\mathfrak{sp}(2)$.
\end{itemize}
We found nondiagonal metrics for all patterns above, and diagonal metrics with the $(10,5)$ pattern only (for both $k$). The second pattern is interesting as we find it is a \textit{pattern breaking} of $(6,9)$, where the 1D eigenspace with corresponding eigenvalue $0.686$ along with the 2D and 3D eigenspaces, form a Lie algebra isomorphic to $\mathfrak{su}(2) \oplus \mathfrak{su}(2)$ (similar to Pauli words with weight one), while the $4\times 4$ skew-Hermitian matrix in the other 1D eigenspace has the eigenvalue structure of a tensor product. For $k=200$, some of solutions with pattern $(1,3,1,4,4,2)$ have that property as well. Finally, the 6D degeneracy for the last pattern does not form a Lie algebra, and the matrix in the 1D eigenspace does not have the eigenvalue structure of a tensor product.

2- $L_{2,6}$ and $\mathfrak{su}(8)$ with $k = 500,1000$:
\begin{itemize}
    \item LM: Only the last pattern occurs exclusively for $k=1000$.
    \begin{enumerate}
        \item $(30,32,1),(0.341, 2.375, 9.901)$
        \item $(10,15,1,32,5), (0.3771,0.3835,0.4432,1.3020, 2.8011)$
        \item $(1,9,1,12,16,9,8,4,3), (0.352, 0.359, 0.381, 0.671,0.689,1.197,1.25,12.5,20.0)$  
    \end{enumerate}
    \item LLM: $(10,15,1,32,5)$ for both $k$.
\end{itemize}
In diagonal solutions, only $(30,32,1)$ appears, while the other two patterns occur in nondiagonal solutions. We know by taking all brackets that the eigenvectors in $1+30$ (and $30$ by itself) give a Lie algebra, but we have not classified it. The same property holds for $(10,15,1,32,5)$ where $1+5+10+15$ (and $5+10+15$) is a $31$D ($30$D) Lie algebra. 

Given that $(9,27,27)$ is the penalty pattern, it can only be related to $(1,9,1,12,16,9,8,4,3)$ if we were to observe some \textit{pattern breaking} similar to the instance in $\mathfrak{su}(4)$. It is the case (up to $1e-3$ errors) that one of the two nines has structure constants similar to that of Pauli words with weight one, i.e. $\mathfrak{su}(2) \oplus \mathfrak{su}(2) \oplus \mathfrak{su}(2)$. However, the matrix in each of the 1D eigenspaces can not be a tensor product of three $2\times 2$ matrices, although they can be a tensor product of a $4\times 4$ and $2\times 2$ matrix. Given that we are only approximating the functional, this suggests that maybe locality does not fully form and one should consider the possibility of partial \textbf{kaq}-ness where there are tensor products of qu\textbf{n}its instead of ($\mathbb{C}^2$) qu\textbf{b}its.

3- $L_{2,4}$ and $\mathfrak{su}(4)$ with $k=100,200$:
\begin{itemize}
    \item LM:
    \begin{enumerate}
        \item $(10,5), (0.790, 1.602)$ only for $k=100$.
        \item $(1,3,1,8,2),(0.347, 0.407, 0.448, 0.971, 10.921) $ for both $k$.
        \item $(6,4,1,4), (0.383, 0.448, 1.645, 4.794)$  only for $k=200$.
        \item $(1,1,2,4,2,1,2,2), (0.365, 0.404, 0.548, 0.586, 0.653, 1.222, 2.118, 9.039)$ only for $k=200$.
    \end{enumerate}
    \item LLM for both $k$: $(1,3,1,8,2)$. 
\end{itemize}
We found diagonal metrics with the $(10,5)$ pattern ($k=100$) and $(6,4,1,4)$ ($k=200$). The latter corresponds to the Majorana fermion degeneracy pattern, where $\psi_1,\ldots,\psi_4$ form a penalty pattern $(4,6,4,1)$, but note the order is not the same. This could be due to the competition the functional feels between forming two local minima, the $\mathfrak{so}(4)$ local minima (same as the penalty pattern $(6,9)$) and the Majorana pattern $(4,6,4,1)$, explaining why $6$ falls before $4$ in the ordered pattern.

The pattern $(1,1,3,8,2)$ is related to the penalty pattern similar to $(3,1,1,8,2)$ in $L_{2,6}$ for $\mathfrak{su}(4)$. The case of $(1,1,2,4,2,1,2,2)$ is yet unknown, however we can show numerically that it is not a pattern breaking of a penalty pattern.

4- $L_{2,4}$ and $\mathfrak{su}(8)$ with $k=1000$:
\begin{itemize}
    \item LM:
    \begin{enumerate}
        \item $(30,32,1), (0.520,  1.620, 61.728)$
        \item $(1,1,15,1,2,4,2,16,16,2,1,2), (0.543, 0.549, 0.552, 0.606, 0.742, 0.746, 0.776, 
        \\0.86, 1.268, 2.994, 3.416, 5.714)$  
    \end{enumerate}
    \item LLM: $(1,1,15,1,2,4,2,16,16,2,1,2)$.
\end{itemize}
Only $(30,32,1)$ is found among diagonal metrics and it is related to a Lie algebra similar to the pattern $(30,32,1)$ in $L_{2,6}$. The second pattern is also related to a similar Lie algebra as $1+1+1+1+2+2+2+2+4+15$ forms a 31D Lie algebra. The 1D eigenspaces are matrices with eigenvalue structure that can be realized by a tensor product ($4\times4$ and $2\times2$). However, none of the patterns can be a breaking of the penalty pattern $(9,27,27)$.
~
\\

\subsection{\textbf{PenMetricOmega} and \textbf{PenMetricPattern} results}~
\\
Using penalty metric intializations, local minima with such patterns are obtained. As noted in Remark \ref{patternequalsomega}, we need to check if the eigenvalues are generated by a weight parameter which is smaller than one.

For $\mathfrak{su}(4)$ and both of $L_{2,4},L_{2,6}$, the answer is always affirmative for the lowest local minima in \textbf{PenMetricPattern}. The loss value attained is $\ge 95\%$ of the lowest loss value found (in LLMs in the previous part), and the weight parameter is pushed more towards zero as $k$ increases.

For $\mathfrak{su}(8)$, there are local minima in \textbf{PenMetricPattern}, with $\ge 95\%$ of the loss value of LLM previously listed, which have the same \textit{ordered} degeneracy pattern as the penalty pattern. However the two ratios $w_2/w_1, w_3/w_2$ do not match; the relative difference is about $30\%$, more than what could be attributed to numerical errors (although it could be due to missing terms in our expansion of the integral). The lowest local minima are given by degeneracy patterns with order $(27+9,27)$ for $L_{2,6}$, where weight one and three have merged, and $(27,27,9)$ with weight two being the smallest for $L_{2,4}$. Finally, The solutions with \textbf{PenMetricOmega} initialization achieved $\sim 90\%$ of the LLM loss value, but none were local minima in the space of all metrics.~
\\
\subsection{Ricci scalar curvature results}~
\\
As mentioned in the introduction, our search for minimizing $||\nabla R||$ yielded the results in \cite{jensen1971scalar}. Thus, only diagonal solutions in $\textbf{sub}_2 \cap \textbf{kaq}_2$ were found corresponding to $\mathfrak{so}(4)$ (6D) and $\mathfrak{sp}(2)$ (10D) Lie subalgebras. 

\section{Summary and Outlook}\label{s4}
There are many directions in which this work can be refined. One could start by investigating the nature of those 31D Lie subalgebras in $\mathfrak{su}(8)$, while collecting more data for \textbf{PenMetricOmega} and more importantly a larger set of (especially higher) values of $k$. Results in this work and preliminary unreported results for $k>200$ for $\mathfrak{su}(4)$, suggest that new patterns emerge for higher values of $k$, which are consolidation of previous ones known for lower $k$s. Perhaps there is a tendency for energy-close eigenspaces to cluster as $k$ increases while the energies of truly distinct eigenspaces grow further apart. We can then ask about the low-energy limit, and what eigenspaces remain.

With regards to \textbf{kaq}-ness, we still do not have a mechanism/theory to prove it given a certain degeneracy pattern, other than checking the Lie subalgebra structure. By doing so, we are ignoring the possibility that a solution is \textbf{kaq} but does not have any Lie subalgebra among (any proper collection of) its eigenspaces. 
One way to address this issue is a gradient descent in the space of all Lie algebra isomorphisms $j^{-1}$, as mentioned in section \ref{s3}.

Also, we did not use any post-processing technique like Pad\'{e} approximates, which could provide a more accurate estimation of the functional.

Another direction is the study of complex fermionic perturbed Gaussian integrals which could reveal similar patterns.

Finally, we would like to have similar numerical simulations for other $\mathfrak{su}(N)$s like $\mathfrak{su}(6)$, to see if \textbf{kaq}-ness would still manifest itself in the form of $\mathbb{C}^3 \otimes \mathbb{C}^2$, showing that there is nothing special about $N=2^n$ regarding decomposition to local tensor factors.

\section*{Acknowledgments}
We would like to thank Carl Bender, Dror Bar-Natan, Scott Morrison, and Xiaoliang Qi for stimulating conversations. This research was done as part of an internship of the second named author at Microsoft, and the writing of this paper was finished while he was supported by Perimeter Institute for Theoretical Physics. Research at Perimeter Institute is supported by the Government of Canada through Innovation, Science and Economic Development Canada and by the Province of Ontario through the Ministry of Research, Innovation and Science.

\appendix
\section{The structure constants can only distinguish weights}
We are interested in exploring the symmetries of the \textit{anti-commutativity} graph of Pauli words $G_{\textbf{PB}_n}$ defined below. 
\begin{dfn}
The graph of Pauli words $G_{\textbf{PB}_n}$ has vertices $\textbf{PB}_n$, and two vertices are connected when they anti-commute.
\end{dfn}

As a simple initial observation, note that all Pauli words have identical spectra, half the eigenvalues are 1 and half -1, thus they are all in the same adjoint orbit, so it turns out to be a rather subtle matter to find asymmetric structures in $G_{\textbf{PB}_n}$ which have any hope of distinguishing, for example, high weight from low weight Pauli words. In this appendix, we initiate the study of such subtle asymmetries in the hope that it will allow us in the future to reverse-engineer functionals which break to penalty metrics.

If there are asymmetries in this graph, especially among the vertices with different weights, one can hope to derive a metric sensitive to such asymmetries. Recall a graph is $k$-ultrahomogeneous if any given isomorphism between two of its induced subgraph with at most $k$ vertices can be extended to an automorphism of the graph. As the following fact demonstrates, $G_{\textbf{PB}_n}$ is symmetric.
\begin{fct}\label{2homogeneity}
$G_{\textbf{PB}_n}$ is a $2^{2n-1}$-regular $2$-ultrahomogeneous graph. The $2$-ultrahomogeneity is facilitated by the Clifford group transformations.
\end{fct}
\begin{rmk}
Following this fact, one has graph automorphisms of $G_{\textbf{PB}_n}$ where vertices of different weights are sent to each other. Moreover, it is a standard Lie algebra fact that the Lie algebra automorphisms of $\mathfrak{u}(N)$ all come from conjugations by $\operatorname{U}(N)$ and a complex conjugation ($\mathbb{Z}_2$ of the Dynkin diagram),  hence by definition of $C_n$, all Lie algebra automorphisms that preserve $\textbf{P}_n$ come from the Clifford group and complex conjugation, and the latter acts trivially on $\textbf{PB}_n$.
\end{rmk}
\begin{rmk}
The fact above also means that a simple preference for more commutativity (and less anti-commutativity or less $|| \ [a,b] \ ||$) can not be the sole reason behind the possible emergence of a penalty metric. Hence minimizing simple functionals like $\sum || [a,b] ||$ will not deliver the desired result.
\end{rmk}
However, we will show that the additional structure imposed by $c_{ij}^k$ on this graph is exactly what is needed to distinguish the vertices with different weight. We recall a fact that will be useful for our next theorem.
\begin{fct}
For every $a,b \in \textbf{P}_n$, either $[a,b] = 0$ or $ab = -ba$. Hence, $c_{ij}^k = 0, \forall k$ or $c_{ij}^k \neq 0 $ for a unique $k$, implying $ij = \frac{c_{ij}^k}{2}k$, and $c_{ij}^k$ satisfies skew- symmetry, i.e. $c_{ik}^j = -c_{ij}^k$, etc.
\end{fct}
We ask if it is possible to have a graph automorphism which preserves the structure constants $c_{ij}^k$ while not preserving weights (see (\ref{penalty_metric}) for the definition of weight)? The negative answer below gives meaning to the title of this section and provides the theoretical motivation for the possibility of penalty metric to emerge from optimization of Gaussian perturbed integrals:
\begin{thm}
Given any $\phi \in \operatorname{Aut}(G_{\textbf{PB}_n})$, $c_{\phi(i)\phi(j)}^{\phi(k)} = c_{ij}^k, \forall i,j,k$ if and only if $\phi$ is a permutation of tensor factors composed with some local automorphism, i.e. $\phi \in \operatorname{Aut}(G_{\textbf{PB}_1})^{\otimes n} \times S_n$, consequently preserving the weight of each word.
\end{thm}
\begin{proof}
In the following, any Pauli word basis element like $-iX\otimes 1\otimes \ldots\otimes 1$ is represented a $X1\ldots1$, where tensors and $-i$ are dropped for convenience. We wish to prove that $\phi(X1\ldots1)$ has weight one. Assuming otherwise, without loss of generality, possibly after composing with a suitable automorphism in $\operatorname{Aut}(G_{\textbf{PB}_1})^{\otimes n} \times S_n$, we have $\phi(X1\ldots1) = X\ldots X 1 \ldots 1$ where there are $l>1$ many $X$s.

Consider the triangle $(i = X1\ldots1,j = Y1\ldots1,k = Z1\ldots1)$ which has $c_{ij}^k = +2$. Sitting at the opposite side of each of $i,j,k$ with respect to the triangle edges, and not connected to them, we have the subgraphs $S_i = X\mathfrak{su}(2^{n-1}), S_j = Y\mathfrak{su}(2^{n-1}), S_k = Z\mathfrak{su}(2^{n-1})$, respectively. Furthermore, every vertex $s_j \in S_j$ is connected to $i$ (and $k$), and has a counterpart $i.s_j = s_k \in S_k$ such that $c_{is_j}^{s_k} = +2$. There exists a similar structure for $j,S_k,S_i$ and $k,S_i,S_j$. Finally, there exists a fourth subgraph $S_{ijk} = 1\mathfrak{su}(2^{n-1})$ not connected to the triangle $ijk$. This gives the structure of the whole $G_{\textbf{PB}_n}$ when viewed from this triangle (see Figure \ref{graphstructure}).

\begin{figure}[h]
    \centering
    \includegraphics[scale = 0.8]{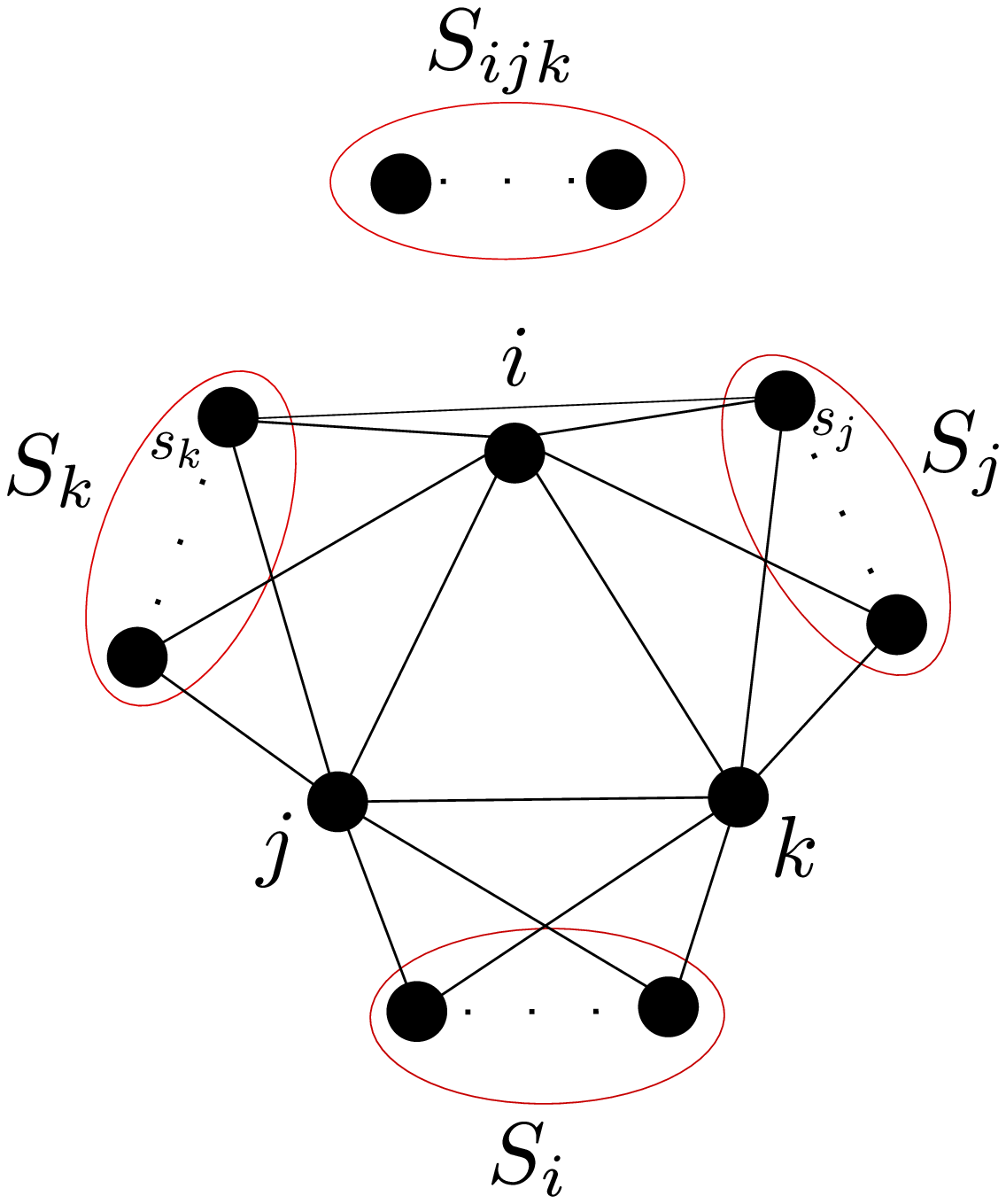}
    \caption{For each triangle with $ij = -ji = k$, one can partition the vertices into four sets $S_i,S_j,S_k,S_{ijk}$, where $i$ is connected to all vertices in $S_j,S_k$ (and similarly for $j,S_i,S_k$ and $k,S_i,S_j$) and $S_{ijk}$ is not connected to any of $i,j,k$. Further, for each $s_j \in S_j$, there is a corresponding vertex $s_k \in S_k$ such that $c_{is_j}^{s_k} \neq 0$ and $is_j = \frac{c_{is_j}^{s_k}}{2} s_k$.}
    \label{graphstructure}
\end{figure}

Using Fact \ref{2homogeneity}, Given any triangle $ijk$ with $c_{ij}^k \neq 0$, the previous description of $G_{\textbf{PB}_n}$ holds for any such $i,j,k$, with the exception of the fact about $c_{is_j}^{s_k}= +2$ for all $s_j,s_k$ corresponding pairs. This means all triangles with $c_{ij}^k \neq 0$ have three subgraphs corresponding to each vertex and not connected to them, and each vertex is connected to the other two subgraphs completely. Also, there is a fourth subgraph not connected to the triangle.

As a result, if $\phi(X1\ldots1) = X\ldots X 1 \ldots 1$, since $\phi$ is a graph automorphism, it must send the triangle and structure described above to that of $ (\phi_x = X\ldots X 1 \ldots 1, \phi_y = \phi(Y1\ldots 1), \phi_z = \phi(Z1\ldots1))$. Therefore, as $\phi$ preserves $c$, we must have $c_{\phi_xs_{\phi_y}}^{s_{\phi_z}} = +2$ for all corresponding pairs $s_{\phi_y},\phi_x.s_{\phi_y}=s_{\phi_z}$. This implies that vertices of $S_{\phi_y}$ are Pauli words where the first $l$ letters have odd many $Y$s and even many $Z$s. However, since $c_{\phi_x\phi_y}^{\phi_z} = +2$, the same is true about $\phi_y$. As $\phi_y$ is not connected to $S_{\phi_y}$, this means the set of Pauli words where the first $l$ letters have odd many $Y$s and even many $Z$s must have an \textit{isolated} vertex. In its first $l>1$ letters,
\begin{itemize}
    \item if $\phi_y$ has at least three $Y$s, then one can exchange two of those with an $X$ and a $1$, to find a neighbor of $\phi_y$ in $S_{\phi_y}$.
    \item If instead $\phi_y$ has one $Y$ in its first $l$ letters, then the whole word $\phi_y$ has either
    \begin{itemize}
        \item two $Z$s in its first $l$ letters, in which case we replace them by $X$ and $1$, or
        \item an $X$ in its first $l$ letters, in which case we replace $Y$ by $1$ and $X$ by $Y$, or
        \item an $X$ or $Y$ or $Z$ outside of its first $l$ letters, in which case we replace it by $Y$, $X$ or $Z$,
    \end{itemize} 
    and in all these cases we find a neighbor in $S_{\phi_y}$.
\end{itemize}
The only remaining case is when $\phi_y = Y1\ldots 1$. Applying the same argument to $\phi_z$ (this time with odd many $Z$s and even many $Y$s) implies $\phi_z = Z1\ldots1$, and since $\phi_x.\phi_y = \phi_z \implies \phi_x = X1\ldots1$.

This proves that $\phi$ has to preserve the weight one Pauli words and modulo some local automorphism and permutation, that $\phi(X1\ldots1) = X1\ldots1$. As $S_{\phi_y}$ must have odd many $Y$s in its first $l=1$ letters and $Y1\ldots1$ is the only isolated vertex among such words, we have $\phi_y = Y1\ldots1$, and similarly $\phi_z = Z1\ldots1$. Restricting $\phi$ to each of the four subgraphs $S_i = X\mathfrak{su}(2^{n-1}), S_j = Y\mathfrak{su}(2^{n-1}), S_k = Z\mathfrak{su}(2^{n-1}),1\mathfrak{su}(2^{n-1})$, the proof follows by induction.
\end{proof}
\begin{rmk}\label{cliffordonc}
Let $\phi \in C_n$. For any $a \in \textbf{PB}_n$, one can write $\phi(a) = \eta_a \phi_a$, for some $\phi_a \in \textbf{PB}_n, \eta_a \in \{\pm 1\}$ (notice $\eta_a \neq \pm i$ as $a, \phi_a$ are skew-Hermitian). Then, if $[a,b] \neq 0$, for some unique $d \in \textbf{PB}_n$:
$$ab = \frac{c_{ab}^d}{2}d \implies \phi(a)\phi(b) = \frac{c_{ab}^d}{2} \phi(d) \implies c_{\phi_a\phi_b}^{\phi_d} = \frac{\eta_c}{\eta_a\eta_b}c_{ab}^d .$$
The theorem above states that there is no $\phi \in C_n$, for which $\frac{\eta_d}{\eta_a\eta_b} = 1$ whenever $c_{ab}^d \neq 0$.
\end{rmk}

\section{Gradient descent and degeneracy patterns}\label{app_preserv_diag}
In this section, we prove Theorem \ref{preservedegeneracypattern}. First, we show that gradient descent preserves diagonal metrics.
\begin{thm}
Given any trivalent tensor network $T(c,g)$ without loops, with vertices labelled by $c$ (structure constants of Pauli word basis), and edges labelled by $g,g,g^{-1}$ for type $i,j,k$ edges, we have
$$\frac{\partial T}{\partial g_{ab}} |_D = 0,$$
for all diagonal metrics $D$ and $a \neq b$.
\end{thm}
\begin{proof}
To compute a tensor contraction, one selects indices at the two ends of each edge, e.g. for an edge of type $i$ the selection $i,i'$ yields a term like $g_{ii'} c_{i-}^{-}c_{i'-}^{-}$.

Thus, differentiating such an expression is done by selecting any edge and choosing labels $a,b$ on the two ends, while the rest of the edges are labelled using the diagonal metric $D$ (see Figure \ref{differentiating} for more details). We wish to show that if $a\neq b$, any such network contracts to zero. To do so, we prove the structure constants multiplier $\prod\limits_{i,j,k \text{ labels on edges}} c_{ij}^{k}$ obtained from any label selection over the edges, is zero if \textit{exactly} one edge is picked nondiagonally, like in $a,b$.

\begin{figure}[h]
    \centering
\begin{tikzpicture}
\Text[x=-2,y=-1]{$\frac{\partial T}{\partial g_{ab}} |_{g=D}=$}
\Vertex[label=$c$]{A}
\Vertex[y=-2,label=$c$]{B}
\Edge[label = $D^{-1}$, color = red](A)(B)
\Edge[bend=65,label=$D$](A)(B)
\Edge[bend=-65,label=$E_{ab}$](A)(B)
\begin{scope}[shift={(3,0)}]
\Text[x=-1.5,y=-1]{$+$}
\Vertex[label=$c$]{A}
\Vertex[y=-2,label=$c$]{B}
\Edge[label = $D^{-1}$, color = red](A)(B)
\Edge[bend=65,label=$E_{ab}$](A)(B)
\Edge[bend=-65,label=$D$](A)(B)
\end{scope}
\begin{scope}[shift={(7,0)}]
\Text[x=-2,y=-1]{$- \frac{\det(D_{a,b})}{\det(D)}$}
\Vertex[label=$c$]{A}
\Vertex[y=-2,label=$c$]{B}
\Edge[label = $E_{ba}$, color = red](A)(B)
\Edge[bend=65,label=$D$](A)(B)
\Edge[bend=-65,label=$D$](A)(B)
\end{scope}
\end{tikzpicture}
    \caption{Differentiating the Theta diagram at a diagonal metric $D$. Differentiating the $i$ or $j$ edges is easy, as it just requires replacing the matrix by the elementary matrix $E_{ab}$. Note the last term requires computing $\frac{\partial g^{-1}}{\partial g_{ab}}$ at $D$. Considering the entries, we need to compute $(-1)^{m+n}\frac{\partial\frac{ \det(g_{mn})}{\det(g)}}{\partial g_{ab}}$, where $g_{mn}$ are the minors. It is straightforward to show that evaluating this expression at a diagonal metric $D$ leaves a nonzero term only for $m=b,n=a$. The matrix $D_{a,b}$ is the diagonal matrix obtained by removing the two $a,b$-th rows and the two $a,b$-th columns.}
    \label{differentiating}
\end{figure}

Without loss of generality, assume that $a,b$ is chosen on an edge of type $i$. Edges of type $j$ and $i$ form a cycle decomposition of the graph. Consider the product of the indices over those edges; as every edge has two equal indices on both ends (except for $a,b$), the product is some scalar times $a.b$. 

On the other hand, if the structure constants multiplier is nonzero, it means at every vertex, $i,j,k$ were chosen such that $c_{ij}^k \neq 0$. Furthermore, in that product, every $i$ is followed by a $j$ giving $\frac{c_{ij}^k}{2}k$. The $k$s are each repeated twice as they appear twice on each edge. As a result, the product should be a scalar, but $a\neq b$, so $a.b$ is not a scalar and we reach contradiction.
\end{proof}
\begin{cor}
Starting at a diagonal metric, gradient descent on loss functions defined in (\ref{L_26},\ref{L_24}) stays in the diagonal metrics space.
\end{cor}
\begin{proof}
Both (\ref{L_26},\ref{L_24}) are linear combinations of tensor networks, and the $(\det(g)-1)^2$ term can be easily checked to have the same property under gradient descent.
\end{proof}
Recall how a subgroup $H<C_n$ defines a degeneracy pattern (\ref{preservedegeneracypattern}).
\begin{thm}\label{app_preserv_diag_thm}
Let $D$ be diagonal with a degeneracy pattern determined according to the orbits of a subgroup $H<C_n$. The gradient flow of loss functions $(\ref{L_26},\ref{L_24})$ preserves not only the diagonal form of the metric, but also the degeneracy pattern modulo the merging of the eigenspaces, e.g $\{d_1,d_2,d_3\}$ could become $\{d_1,d_2+d_3\}$.
\end{thm}
\begin{proof}
We previously showed that $D$ stays diagonal under gradient flow. Next, we need to show that entries corresponding to the same eigenspace stay equal, i.e. have equal gradients. It suffices to prove so for every tensor network $T$; as the determinant term $(\det(g)-1)^2$ has the same property, the statement would follow.

Assume $g_{aa} = g_{bb}$ where there is an $h\in H$ such that $h(a) = \eta_a b$ where $\eta_a \in \{\pm 1\}$. To show $\frac{\partial T}{\partial g_{aa}} = \frac{\partial T}{\partial g_{bb}}$, we use the same picture as in the previous theorem. Every tensor contraction in the summation giving $\frac{\partial T}{\partial g_{aa}}$ is given by fixing both ends of some edge $e$ with label $a$. This tensor is sent to an equal contraction in $\frac{\partial T}{\partial g_{bb}}$ by simply applying $h$ on the tensor. Indeed, since $g_{xx}=g_{h(x)h(x)}$ for any $x$, the coefficients involving $g$ entries are equal. As for the structure constants multiplier,  since $h(i) = \eta_i h_i$ for some $h_i$ in the Pauli word basis, we need to show that $$\prod\limits_{i,j,k \text{ selected labels}} c_{ij}^{k} = \prod\limits_{i,j,k \text{ selected labels}} c_{h_ih_j}^{h_k}.$$
This is true for \textit{any} $h \in C_n$. Indeed, as computed in Remark \ref{cliffordonc},
$$c_{h_ih_j}^{h_k} = \frac{\eta_k}{\eta_i\eta_j}c_{ij}^k,$$
and since in the product $\prod\limits_{i,j,k \text{ selected labels}} c_{ij}^{k}$ every label appears twice and  $\eta_{x} \in \{\pm1\},\forall x$, we are left with the equality of the two products.
\end{proof}
\begin{rmk}
While on one hand, this theorem implies that one can find local minima with a penalty degeneracy pattern by simply initializing at such a pattern, it also means that the penalty pattern is not special in that regard, as we can find local minima with many other degeneracy patterns.
\end{rmk}

\section{Individual diagrams are convex}\label{tensorsconvex}
Experiments show that the individual tensor networks used in the asymptotic expansion of (\ref{theintegral}) are all positive convex functions with unique minimum at $g=\operatorname{Id}$.  We can provide some theoretical evidence in this regard for some of the diagrams.

We start with the Theta diagram $\Theta(c,g)$. Assume $g=UDU^\dagger$ where $D$ is diagonal and $U$ unitary. We wish to show $\min_D \Theta(c,UDU^\dagger) = \Theta(c,\operatorname{Id})$.

We can rewrite the expression above as $\min_D \Theta(c_U,D) = \Theta(c_U,\operatorname{Id}) = \Theta(c,\operatorname{Id})$, where the edges of tensor $c$ with the tensors $U$ (or $U^\dagger$) are contracted. We have
$$\Theta(c_U,D) = \sum_{i,j,k} |c_{U(i)U(j)}^{U(k)}|^2 d_i d_j d_k^{-1}.$$
Writing $d_i = e^{w_i}$, $\Theta(c_U,D)$ becomes a positive sum of exponentials of affine functions in $w_i$, hence convex. By using the arithmetic-geometric mean inequality, the minimum can be seen to occur at $g=\operatorname{Id}$. As previously mentioned, experiments have shown that this minimum is unique.

We can prove similar results for other diagrams such as the Tincan, by bringing the contraction in the form written above for $\Theta(c_U,D)$, through successive minimization over some edges, while others are assumed fixed and once contracted give two equal tensors being contracted over the same indices, same as in $\Theta$.

For the Tincan $T(c,g)$, the starting idea is similar, fixing $U$ in the diagonalization of $g$ and letting $D$ be the variable. However, at first, the vertical edges are also assumed to be fixed, meaning the variable is $D$ on the $4$ vertical edges. The contraction of tensors along the vertical lines gives two $4$-valent tensors $A,B$ connected with $D$ (or $D^{-1}$) along each edge. The two tensors are equal due to the symmetry of the diagram and the previous contraction being done along the same index. Hence, we obtain an expression similar to the one for $\Theta(c_U,D)$, implying the minimum is attained at $D=\operatorname{Id}$ on the horizontal edges, given the vertical edges. Next step is to consider $D$ on the vertical edges as the variable and the proof follows similarly.

\bibliographystyle{apa}
\bibliography{main}

\bigskip
\address{\textsuperscript{*\label{1}}
	Microsoft Research, Station Q, and Department of Mathematics, University of California, Santa Barbara, CA 93106, USA 
}

\address{\textsuperscript{$\dagger$\label{2}}
	Perimeter Institute for Theoretical Physics, Waterloo, ON N2L 2Y5, Canada 
}

\end{document}